\def\E{\end{document}}
\documentclass[11pt]{article}
\usepackage{amssymb,amsmath}

\begin{document}
\title{
Spacetime estimates and scattering theory for quasilinear Schr\"{o}dinger equations
in arbitrary space dimension}
  \author{Xianfa Song{\thanks{E-mail: songxianfa2004@163.com(X.F. Song)
 }}\\
\small Department of Mathematics, School of Mathematics, Tianjin University,\\
\small Tianjin, 300072, P. R. China
}

\maketitle
\date{}

\newtheorem{theorem}{Theorem}[section]
\newtheorem{definition}{Definition}[section]
\newtheorem{lemma}{Lemma}[section]
\newtheorem{proposition}{Proposition}[section]
\newtheorem{corollary}{Corollary}[section]
\newtheorem{remark}{Remark}[section]
\renewcommand{\theequation}{\thesection.\arabic{equation}}
\catcode`@=11 \@addtoreset{equation}{section} \catcode`@=12

\begin{abstract}

In this paper, we consider the following Cauchy problem of
\begin{equation*}
\left\{
\begin{array}{lll}
iu_t=\Delta u+2\delta_huh'(|u|^2)\Delta h(|u|^2)+V(x)u+F(|u|^2)u+(W*|u|^2)u,\ x\in \mathbb{R}^N,\ t>0\\
u(x,0)=u_0(x),\quad x\in \mathbb{R}^N.
\end{array}\right.
\end{equation*}
Here $\delta_h$ is a constant, $N\geq 1$, $h(s)$, $F(s)$, $V(x)$ and $W(x)$ are some real functions, $W(x)$ is even. Besides obtaining some sufficient conditions on global existence of the solution, we
 establish pseudoconformal conservation law and give Morawetz type estimates, spacetime bounds and asymptotic behaviors for the global solution.

 We bring two ideas to establish scattering theory, one is that we take different admissible pairs in Strichartz estimates for different terms on the right side of Duhamel's formula in order to keep each term independent, another is that we factitiously let a continuous function be the sum of two piecewise functions and chose different admissible pairs in Strichartz estimates for the terms containing these functions.
 Basing on the two ideas, we provide the direct and simple proofs of classic scattering theories in $L^2(\mathbb{R}^N)$ and $\Sigma$ for any space dimension($N\geq 1$) under certain assumptions. Here
 $$
 \Sigma=\{u\in H^1(\mathbb{R}^N),\quad |xu|\in L^2(\mathbb{R}^N)\}.
 $$

{\bf Keywords:} Qusilinear Schr\"{o}dinger equation; Spacetime estimate; Srtichartz estimate; Scattering.

{\bf 2000 MSC: 35Q55.}

\end{abstract}

\newpage

{\bf Contents.}

1. Introduction.

2. Global existence and pseudoconformal conservation law for the solution of (\ref{1}).

3. Morawetz type estimates based on pseudoconformal conservation law.

3.1. The proof of Theorem 3 in Case 1.

3.2. The proof of Theorem 3 in Case 2.

4. Spacetime estimates based on pseudoconformal conservation law.

5. Interaction Morawetz inequality.

5.1. Interaction Morawetz inequality in dimension $N\geq 3$.

5.2. Interaction Morawetz inequality in dimension $N=2$.

5.3. Interaction Morawetz inequality in dimension $N=1$.

6. Classic scattering theory for (\ref{semilinear1}) in defocusing case and arbitrary space dimension.

6.1. Classic scattering theory in $L^2(\mathbb{R}^N)$ for (\ref{semilinear1}) in defocusing case and arbitrary space dimension.

6.2. Classic scattering theory in $\Sigma$ for (\ref{semilinear1}) in defocusing case and arbitrary space dimension.

References.
\newpage

\section{Introduction}
\qquad In this paper, we consider the following Cauchy problem:
\begin{equation}
\label{1} \left\{
\begin{array}{lll}
iu_t=\Delta u+2\delta_huh'(|u|^2)\Delta h(|u|^2)+V(x)u+F(|u|^2)u+(W*|u|^2)u,\ x\in \mathbb{R}^N,\ t>0\\
u(x,0)=u_0(x),\quad x\in \mathbb{R}^N.
\end{array}\right.
\end{equation}
Here $h(s)$, $F(s)$, $V(x)$ and $W(x)$ are some real functions, $W(x)$ is even, $N\geq 1$.
\begin{equation}
\delta_h=\left\{
\begin{array}{lll}
0,\quad h(s)\equiv 0\\
1,\quad h(s)\geq 0,\ \not \equiv 0.
\end{array}\right.
\end{equation}
(\ref{1}) can be used to model a lot of physical phenomena, such as the superfluid film  equation in plasma physics if $h(s)=s$, physics phenomenon in dissipative quantum mechanics if $h(s)=\sqrt{s}$ and the self-channelling of a high-power ultra short laser in matter if $h(s)=\sqrt{1+s}$. It also appears in condensed matter theory and nonlinear optical theory, see \cite{Bass, Borovskii, Bouard, Goldman, Ku, LSS, Litvak, Makhankov, Ritchie}. There are many interesting topics on (\ref{1}), such as local wellposedness, global wellposeness, decay rate and scattering phenomenon for the global solution.

We would like to say something about the local wellposedness of the solution to (\ref{1}). In convenience, we always assume that $h(s)\geq 0$ for $s\geq 0$, $V(x)\leq 0$ and $W(x)\leq 0$ for $x\in\mathbb{R}^N$ in this paper. We say that (\ref{1}) is in defocusing case if $F(s)\leq 0$ for $s\geq 0$, while we say that (\ref{1}) is in combined defocusing and focusing case if $F(s)\geq 0$ for $s\geq 0$ or changes sign. The assumptions on $V(x)$ and $W(x)$ are as follows:

({\bf WV1})\quad  If $h(s)\equiv 0$ for $s\geq 0$, we require that $V(x)\in L^{p_1}(\mathbb{R}^N)+L^{\infty}(\mathbb{R}^N)$ for some $p_1>\max(1,\frac{N}{2})$ and $W(x)\in L^{p_2}(\mathbb{R}^N)+L^{\infty}(\mathbb{R}^N)$ for some $p_2>\max(1,\frac{N}{4})$

or

({\bf WV2})\quad If $h(s)\geq 0, \not \equiv 0$ for $s\geq 0$, we require that $V(x)\in L^{\infty}(\mathbb{R}^N)$, and $W(x)\in L^1(\mathbb{R}^N)\cap \{L^{p_2}(\mathbb{R}^N)+L^{\infty}(\mathbb{R}^N)\}$ for some $p_2>\max(1,\frac{N}{4})$.

Besides the assumptions on $V(x)$ and $W(x)$, under certain conditions on $F(s)$, (\ref{1}) is local wellposedness in
\begin{align}
X=\{w\in H^1(\mathbb{R}^N),\quad \int_{\mathbb{R}^N}|\nabla h(|w|^2)|^2dx<+\infty\}\label{kongjianziji}
\end{align}
by the results of \cite{Cazenave, Colin1, Colin2, Kenig1, Poppenberg1}.

The asymptotic behavior and scattering phenomenon are very important and interesting topics on the study of nonlinear Sch\"{o}dinger equation.
Pseudoconformal conservation law is essential for the study of the asymptotic behavior for the solution,  Morawetz estimate is an important tool to construct scattering operator on the energy space, see \cite{Bez, Cazenave, Colliander1, Colliander2, Ginibre1, Ginibre2, Ginibre3, Ginibre4, Morawetz,Nakanishi, Ozawa}.

However, two more interesting questions are as follows: 1. What is the relationship between pseudoconformal conservation law and Morawetz estimate? 2. How to establish the link between pseudoconformal conservation law and spacetime bound estimate?

The first motivation of this paper is to obtain the answers of the two questions above. To do this, we will establish Morawetz type estimates and weighted spacetime bounds based on pseudoconformal conservation law in this paper, which reveals the relationship among pseudoconformal conservation law, Morawetz type estimates and spacetime bounds. These results are also very interesting discover in the study of quasilinear Sch\"{o}dinger equation in the following sense: To our best knowledge, although we obtained some results on the asymptotic behaviors for the solution of a quasilinear Schr\"{o}dinger equation containing Hartree type nonlinearity in \cite{SongWang2}, and the related results on Morawetz estimates and weighted spacetime bounds for the solution of a quasilinear Schr\"{o}dinger equation in \cite{Song1, Song2}, there are few results on  Morawetz estimates and weighted spacetime bounds for the solution of (\ref{1}) which contains more general nonlinearities.

The second motivation of this paper is to show some applications of spacetime estimates for the global solution. To do this, one thing is to consider the asymptotic behavior for the solution of (\ref{1}) as $t\rightarrow +\infty$, another one is to establish scattering theory for (\ref{1}) in the case of (WV1), i.e.,
\begin{equation}
\label{semilinear1} \left\{
\begin{array}{lll}
iu_t=\Delta u+V(x)u+F(|u|^2)u+(W*|u|^2)u,\ x\in \mathbb{R}^N,\ t>0\\
u(x,0)=u_0(x),\quad x\in \mathbb{R}^N.
\end{array}\right.
\end{equation}
Many authors obtained scattering results on (\ref{semilinear1}) when at least one of $V(x)\equiv 0$, $F(|u|^2)u \equiv 0$ and $W(x) \equiv 0$ holds. We can refer to \cite{Banica, Bourgain, Carles03, Carles, Cazenave, Colliander2, Colliander3, Costin, Dodson, Duyckaerts, Germain, Ginibre1, Ginibre3, Hayashi, Kenig2, Killip, Killip2, Lu, Nakanishi, Strauss, Tao1, Tao, Tsutsumi1, Tsutsumi2,Visan, Zhang} and the references therein. Especially, in Chapter 7 of the book \cite{Cazenave},  Cazenave introduced systematically the scattering results on the Cauchy problem of $iu_t=\Delta u+|u|^{\alpha}u$. However, to our best knowledge, there are few scattering results on the following special case of (\ref{semilinear1})
\begin{equation}
\label{224x2} \left\{
\begin{array}{lll}
iu_t=\Delta u-\frac{a}{|x|^m}u-b|u|^{2\beta}u-(\frac{c}{|x|^n}*|u|^2)u,\quad x\in \mathbb{R}^N\setminus\{\mathbf{0}\},\ t>0\\
u(x,0)=u_0(x),\quad x\in \mathbb{R}^N,
\end{array}\right.
\end{equation}
$a\neq 0$, $b\neq 0$ and $c\neq 0$, let alone in the general case of $V(x)\not \equiv 0$, $F(|u|^2)u \not \equiv 0$ and $W(x)\not \equiv 0$.

Since we will establish several theorems and the length of formulae are long, we don't state the precise expressions of these theorems in the introduction. To control the length of Section 1, we will state and prove them in the corresponding sections.

However, we would like to say something about them roughly below.

1. About the conditions on global existence of solution to (\ref{1}), if $h(s)\neq 0$, $F(s)=F_1(s)-F_2(s)$ in the combined defocusing and focusing case, $F_1(s)\geq 0$ and $F_2(s)\geq 0$ for $s\geq 0$, $G_1(s)=\int_0^sF_1(\eta)d\eta$, then a criterion is to find
$$0<\gamma<1,\quad \gamma'>1\quad {\rm satisfying}\quad \frac{2^*(1-\gamma)}{2(\gamma'-\gamma)}\leq 1$$
 such that
$$
[|G_1(s)|]^{\gamma}\leq c_1s ,\quad [|G_1(s)|]^{\gamma'}\leq c'_1[s^{\frac{1}{2}}+h(s)]^{2^*}.
$$

2. We will establish pseoduconformal conservation law, which is essential for the study of the asymptotic behavior for the global solution of (\ref{1}). Basing on it, we give Morawetz type estimates, which reveals the relationship between pseoduconformal conservation law and Morawetz estimate.

3. About the decay rate of the solution to (\ref{1}), we obtain
$$
\int_{\mathbb{R}^N}[\delta_h|\nabla h(|u|^2)|^2+|V(x)||u|^2+|G_1(|u|^2)|+|G_2(|u|^2)+\frac{1}{2}(|W|*|u|^2)|u|^2]dx\leq \frac{C}{t^{\iota}}
$$
for some $0<\iota\leq 2$ and asymptotic behavior
$$
|\int_{\mathbb{R}^N}|\nabla u(x,t)|^2dx-2E(u_0)|\leq \frac{C}{t^{\iota}},\quad \lim_{t\rightarrow +\infty}\int_{\mathbb{R}^N}|\nabla u|^2dx=2E(u_0).
$$
under certain conditions.

4. Under certain assumptions, we establish Morawetz type estimates such as
\begin{align*}
&\quad\int_0^{+\infty}\int_{\mathbb{R}^N}\frac{\left[\delta_h|\nabla h(|u|^2)|^2+|V(x)||u|^2+|G_1(|u|^2)|+|G_2(|u|^2)|+\frac{1}{2}(|W|*|u|^2)|u|^2\right]^{\theta}}{a(x,t)}dxdt\nonumber\\
&\leq M_1(u_0,\theta),
\end{align*}
and weighted spacetime bounds such as
\begin{align*}
\|G_1(|u|^2)\|_{L_w^q(\mathbb{R}^+)L_w^r(\mathbb{R}^N)}&=
\left(\int_0^{+\infty}\left(\int_{\mathbb{R}^N}w(x,t)[|G_1(|u|^2)|]^rdx\right)^{\frac{q}{r}}dt\right)^{\frac{1}{q}}\leq C(u_0,r,q).
\end{align*}

5. Under certain assumptions, we establish classic scattering theory for (\ref{semilinear1}) with general $V(x)$, $F(|u|^2)$ and $W(x)$,
$$\|e^{it\Delta}u(t)-u_+\|_{L^2}\rightarrow 0\quad {\rm as}\quad t\rightarrow +\infty.$$
Especially, if $V(x)\equiv 0$ and $W(x)\equiv 0$, we can obtain the scattering result on (\ref{semilinear1}) with general $F(|u|^2)$,
$$\|e^{it\Delta}u(t)-u_+\|_{\Sigma}\rightarrow 0\quad {\rm as}\quad t\rightarrow +\infty.$$

It is the idea to keep each term independent that: We take different admissible pairs in Strichartz estimates for different terms on the right side of Duhamel's formula in the proof.

Another idea is to factitiously let a continuous function be the sum of two piecewise functions and chose different admissible pairs in Strichartz estimates for the two terms. For example,
let $\frac{1}{|x|^m}=V_1(x)+V_2(x)$, where
\begin{equation*}
V_1(x)=\left\{
\begin{array}{lll}
\frac{1}{|x|^m},\quad 0<|x|\leq 1,\\
0,\quad |x|>1,
\end{array}\right.\quad {\rm and}\quad
V_2(x)=\left\{
\begin{array}{lll}
0,\quad 0<|x|\leq 1,\\
\frac{1}{|x|^m},\quad |x|>1.
\end{array}\right.
\end{equation*}
Then we have
\begin{align}
& \|\int_t^{\tau} e^{is\Delta}[\frac{1}{|x|^m}u(s)]ds\|_{L^2}=\|\int_t^{\tau} e^{is\Delta}[V_1(x)u(s)+V_2(x)u(s)]ds\|_{L^2}\nonumber\\
&\leq \|\int_t^{\tau} e^{is\Delta}V_1(x)u(s)ds\|_{L^2}+\|\int_t^{\tau} e^{is\Delta}V_2(x)u(s)ds\|_{L^2}\nonumber\\
&\leq C\sum_{j=1}^2\left(\int_t^{\tau}\left(\int_{\mathbb{R}^N}|V_j(x)u|^{r'_j}dx\right)^{\frac{q'_j}{r'_j}}dt\right)^{\frac{1}{q'_j}}.\label{812w1}
\end{align}
Here $(q_j, r_j)$, $j=1,2$, are admissible pairs, $q'_j$ and $r'_j$ are the conjugated exponents of $q_j$ and $r_j$ respectively.
We believe that the two ideas can also be applied to study scattering phenomenon on the related problems of Schr\"{o}dinger equations.

The organization of this paper is as follows. In Section 2, we will prove mass and energy conservation laws, obtain some sufficient conditions on the global existence of the solution to (\ref{1}) and establish pseudoconformal conservation law.
In Section 3, we will give Morawetz type estimates based on pseudoconformal conservation law. In Section 4, we consider spacetime bound estimates for the solution. In Section 5, we will give interaction Morawetz estimates for the solution. In Section 6, we will establish classic scattering theory for (\ref{semilinear1}) as the applications of these estimates.

\section{Global existence and pseudoconformal conservation law for the solution of (\ref{1})}
\qquad In convenience, we will use $C$, $C'$, and so on, to denote some constants in the sequels, the values of it may vary line to line.

In this section,  we will prove mass and energy conservation laws, obtain some sufficient conditions on the global existence of the solution to (\ref{1}) and establish pseudoconformal conservation law.

First, we prove a lemma as follows.

{\bf Lemma 2.1.} {\it Assume that $u$ is the solution of (\ref{1}). Then in the time interval $[0,t]$ when it exists, $u$ satisfies

(i) Mass conversation: $$ m(u)=\left(\int_{\mathbb{R}^N}|u(x,t)|^2dx\right)^{\frac{1}{2}}=\left(\int_{\mathbb{R}^N}|u_0(x)|^2dx\right)^{\frac{1}{2}}=m(u_0);$$

(ii) Energy conversation:
\begin{align}
E(u)&=\frac{1}{2}\int_{\mathbb{R}^N}[|\nabla u|^2+\delta_h|\nabla h(|u|^2)|^2]dx-\frac{1}{2}\int_{\mathbb{R}^N}V(x)|u|^2dx\nonumber\\
&\quad -\frac{1}{2}\int_{\mathbb{R}^N}G(|u|^2)dx-\frac{1}{4}\int_{\mathbb{R}^N}(W*|u|^2)|u|^2dx\nonumber\\
&=E(u_0);\label{2227}
\end{align}

(iii) $$\frac{d}{dt} \int_{\mathbb{R}^N}|x|^2|u|^2dx=-4\Im \int_{\mathbb{R}^N} \bar{u}(x\cdot \nabla u)dx;$$

(iv) \begin{align}
&\quad \frac{d}{dt} \Im \int_{\mathbb{R}^N} \bar{u}(x\cdot \nabla u)dx\nonumber\\
 &=-2\int_{\mathbb{R}^N}|\nabla u|^2dx-(N+2)\delta_h\int_{\mathbb{R}^N}|\nabla h(|u|^2)|^2dx\nonumber\\
&\quad -8N\delta_h\int_{\mathbb{R}^N}h''(|u|^2)h'(|u|^2)|u|^4|\nabla u|^2dx-\int_{\mathbb{R}^N}(x\cdot \nabla V)|u|^2dx\nonumber\\
&\quad +N\int_{\mathbb{R}^N}[|u|^2F(|u|^2)-G(|u|^2)]dx-\int_{\mathbb{R}^N}[\frac{(x\cdot \nabla W)}{2}*|u|^2]|u|^2dx.\label{10131}
\end{align}
}

{\bf Proof:} (i) Multiplying (\ref{1}) by $2\bar{u}$, taking the imaginary part of the result, we get
\begin{align}
\frac{\partial }{\partial t}|u|^2=\Im(2\bar{u}\Delta u) =\nabla \cdot (2\Im \bar{u}\nabla u).\label{10121}
\end{align}
Integrating (\ref{10121}) over $\mathbb{R}^N\times [0,t]$, we have
$$ \int_{\mathbb{R}^N}|u|^2dx=\int_{\mathbb{R}^N}|u_0|^2dx,$$
which implies mass conservation law.

(ii)  Multiplying (\ref{1}) by $2\bar{u}_t$, taking the real part of the result, then integrating it over $\mathbb{R}^N\times [0,t]$, we obtain
\begin{align*}
&\quad\int_{\mathbb{R}^N}[|\nabla u|^2+\delta_h|\nabla h(|u|^2)|^2]dx-\int_{\mathbb{R}^N}V(x)|u|^2dx\\
&\quad -\int_{\mathbb{R}^N}G(|u|^2)dx-\frac{1}{2}\int_{\mathbb{R}^N}(W*|u|^2)|u|^2dx\nonumber\\
&=\int_{\mathbb{R}^N}[|\nabla u_0|^2+\delta_h|\nabla h(|u_0|^2)|^2]dx-\int_{\mathbb{R}^N}V(x)|u_0|^2dx\nonumber\\
&\quad -\int_{\mathbb{R}^N}G(|u_0|^2)dx-\frac{1}{2}\int_{\mathbb{R}^N}(W*|u_0|^2)|u|^2dx,
\end{align*}
which implies energy conservation law.

(iii) Multiplying (\ref{10121}) by $|x|^2$ and integrating it over $\mathbb{R}^N$, we get
\begin{align*}
\frac{d}{dt}\int_{\mathbb{R}^N}|x|^2|u|^2dx&=\int_{\mathbb{R}^N}|x|^2\nabla \cdot(2\Im (\bar{u}\nabla u))dx
=-4\Im \int_{\mathbb{R}^N}\bar{u}(x\cdot \nabla u)dx.
\end{align*}

(iv) Let  $a(x,t)=\Re u(x,t)$ and $b(x,t)=\Im u(x,t)$. Then
\begin{align*}
&\quad\frac{d}{dt}\Im \int_{\mathbb{R}^N}\bar{u}(x\cdot \nabla u)dx\nonumber\\
&= \int_{\mathbb{R}^N}\sum_{k=1}^N[x_k(b_t)_{x_k}a-x_k(a_t)_{x_k}b]dx+\int_{\mathbb{R}^N}\sum_{k=1}^N(x_kb_{x_k}a_t-x_ka_{x_k}b_t)dx\nonumber\displaybreak\\
 &=\int_{\mathbb{R}^N} \sum_{k=1}^N[x_k(b_t)_{x_k}a-x_k(a_t)_{x_k}b]dx
+\int_{\mathbb{R}^N} \sum_{k=1}^N (x_ka_{x_k}\Delta a+x_kb_{x_k}\Delta b)dx\nonumber\\
&\ +\frac{1}{2}\int_{\mathbb{R}^N}\sum_{k=1}^N x_k(|u|^2)_{x_k}[2\delta_hh'(|u|^2)\Delta h(|u|^2)+V(x)+F(|u|^2)+(W*|u|^2)]dx\nonumber\\
&=N\int_{\mathbb{R}^N}(a_tb-ab_t)dx+\int_{\mathbb{R}^N}\sum_{k=1}^N(x_kb_{x_k}a_t-x_ka_{x_k}b_t)dx\nonumber\\
&\quad+\frac{N-2}{2}\int_{\mathbb{R}^N}|\nabla u|^2dx+\frac{N-2}{2}\int_{\mathbb{R}^N}\delta_h|\nabla h(|u|^2)|^2dx\nonumber\\
&\quad-\frac{1}{2}\int_{\mathbb{R}^N}[NV+(x\cdot \nabla V)]|u|^2dx-\frac{N}{2}\int_{\mathbb{R}^N}G(|u|^2)dx\nonumber\\
&\quad -\frac{1}{2}\int_{\mathbb{R}^N}([NW+\frac{1}{2}(x\cdot \nabla W)]*|u|^2)|u|^2dx\nonumber\\
&=N\int_{\mathbb{R}^N}\left([a\Delta a+b\Delta b]+2\delta_h|u|^2h'(|u|^2)\Delta h(|u|^2)\right)dx\nonumber\\
&\quad+N\int_{\mathbb{R}^N}\left(V(x)|u|^2+F(|u|^2)|u|^2+(W*|u|^2)|u|^2\right)dx\nonumber\\
&\quad +(N-2)\int_{\mathbb{R}^N}|\nabla u|^2dx+(N-2)\int_{\mathbb{R}^N}\delta_h|\nabla h(|u|^2)|^2dx\nonumber\\
&\quad-\int_{\mathbb{R}^N}[NV+(x\cdot \nabla V)]|u|^2dx-N\int_{\mathbb{R}^N}G(|u|^2)dx\nonumber\\
&\quad -\int_{\mathbb{R}^N}([NW+\frac{1}{2}(x\cdot \nabla W)]*|u|^2)|u|^2dx\nonumber\\
&=-2\int_{\mathbb{R}^N}|\nabla u|^2dx-(N+2)\int_{\mathbb{R}^N}\delta_h|\nabla h(|u|^2)|^2dx\nonumber\\
&\quad-8N\int_{\mathbb{R}^N}\delta_h h'(|u|^2)h''(|u|^2)|u|^4|\nabla u|^2dx-\int_{\mathbb{R}^N}(x\cdot \nabla V)|u|^2dx\nonumber\\
&\quad +N\int_{\mathbb{R}^N}[|u|^2F(|u|^2)-G(|u|^2)]dx-\int_{\mathbb{R}^N}[\frac{(x\cdot \nabla W)}{2}*|u|^2]|u|^2dx.
\end{align*}
Lemma 2.1 is proved.\hfill $\Box$

Next, we establish some sufficient conditions on the global existence of the solution.

{\bf Theorem 1. } {\it Let $u(x,t)$ be the solution of (\ref{1}) with $u_0\in X$. Assume that $V(x)\leq 0$, $W(x)\leq 0$ for $x\in \mathbb{R}^N$, and satisfy (WV1) or (WV2). Then $u(x,t)$ is global existence in one of the following cases:

Case 1.  defocusing case, i.e., $F(s)=F_2(s)\leq 0$ for $s\geq 0$, $N\geq 1$, and the initial data $u_0$ satisfies $0<M(u_0)<+\infty$ and $0\leq E(u_0)<+\infty$;

Case 2.  $h(s)\neq 0$, $F(s)=F_1(s)-F_2(s)$ in the combined defocusing and focusing case, $F_1(s)\geq 0$ and $F_2(s)\geq 0$ for $s\geq 0$, $N\geq 3$, and there exist $c_1, c'_1, c_2, c'_2>0$, $0<\gamma_1, \tilde{\gamma}_1<1$ and $\gamma_2,\tilde{\gamma}_2>1$ such that
\begin{align}
& \frac{2^*(1-\gamma_1)}{2(\gamma_2-\gamma_1)}\leq 1,\quad \frac{2^*(1-\tilde{\gamma}_1)}{2(\tilde{\gamma}_2-\tilde{\gamma}_1)}\leq 1,\label{6261}\\
& [|G_1(s)|]^{\gamma_1}\leq c_1s ,\quad [|G_1(s)|]^{\gamma_2}\leq c'_1[s^{\frac{1}{2}}+h(s)]^{2^*}\ {\rm for}\ 0\leq s\leq 1,\label{6262}\\
& [|G_1(s)|]^{\tilde{\gamma}_1}\leq c_2s ,\quad [|G_1(s)|]^{\tilde{\gamma}_2}\leq c'_2[ s^{\frac{1}{2}}+h(s)]^{2^*}\ {\rm for}\  s>1,\label{6263}
\end{align}
besides $0\leq E(u_0)<+\infty$, the initial data $u_0$ satisfies
$$
\frac{2^*(1-\gamma_1)}{2(\gamma_2-\gamma_1)}=1,\quad \frac{2^*(1-\tilde{\gamma}_1)}{2(\tilde{\gamma}_2-\tilde{\gamma}_1)}<1,\quad (c_1\|u_0\|^2_{L^2})^{\frac{2}{N}}(2^{2^*-1}c'_1C_s)^{\frac{N-2}{N}}<\frac{1}{4},
$$
or
$$
\frac{2^*(1-\gamma_1)}{2(\gamma_2-\gamma_1)}<1,\quad \frac{2^*(1-\tilde{\gamma}_1)}{2(\tilde{\gamma}_2-\tilde{\gamma}_1)}=1,\quad (c_2\|u_0\|^2_{L^2})^{\frac{2}{N}}(2^{2^*-1}c'_2C_s)^{\frac{N-2}{N}}<\frac{1}{4},
$$
or
$$
\frac{2^*(1-\gamma_1)}{2(\gamma_2-\gamma_1)}=1,\quad \frac{2^*(1-\tilde{\gamma}_1)}{2(\tilde{\gamma}_2-\tilde{\gamma}_1)}=1,
\quad\sum_{j=1}^2(c_j\|u_0\|^2_{L^2})^{\frac{2}{N}}(2^{2^*-1}c'_jC_s)^{\frac{N-2}{N}}<\frac{1}{4}.
$$
Here $G_i(s)=\int_0^sF_i(\eta)d\eta(i=1,2)$ and $G(s)=G_1(s)-G_2(s)$, $C_s$ denotes the best constant in the Sobolev's inequality
\begin{align}
\int_{\mathbb{R}^N}w^{2^*}dx\leq C_s\left(\int_{\mathbb{R}^N}|\nabla w|^2dx\right)^{\frac{2^*}{2}}\quad {\rm for \ any}\ w\in H^1(\mathbb{R^N}).\label{zjcs}
\end{align}
}

{\bf The proof of Theorem 1:}

{\bf Case 1.} By energy conservation law and the assumptions on $V(x)$, $W(x)$, we have
\begin{align}
E(u)&=\frac{1}{2}\int_{\mathbb{R}^N}[|\nabla u|^2+\delta_h|\nabla h(|u|^2)|^2]dx+\frac{1}{2}\int_{\mathbb{R}^N}|V(x)||u|^2dx\nonumber\\
&\quad +\frac{1}{2}\int_{\mathbb{R}^N}|G(|u|^2)|dx+\frac{1}{4}\int_{\mathbb{R}^N}(|W|*|u|^2)|u|^2dx\nonumber\\
&=E(u_0)<+\infty.
\end{align}

{\bf Case 2.} Note the fact
\begin{align}
\int_{\mathbb{R}^N}|G_1(|u|^2)|dx&=\int_{\{|u|\leq 1\}}|G_1(|u|^2)|dx+\int_{\{|u|>1\}}|G_1(|u|^2)|dx\nonumber\\
&\leq \left(\int_{\{|u|\leq 1\}}|G_1(|u|^2)|^{\gamma_1}dx \right)^{\frac{1}{\tilde{\tau}'_1}}\left(\int_{\{|u|\leq 1\}}|G_1(|u|^2)|^{\gamma_2}dx \right)^{\frac{1}{\tilde{\tau}_1}}\nonumber\\
&\quad+\left(\int_{\{|u|>1\}}|G_1(|u|^2)|^{\tilde{\gamma}_1}dx \right)^{\frac{1}{\tilde{\tau}'_2}}\left(\int_{\{|u|>1\}}|G_1(|u|^2)|^{\tilde{\gamma}_2}dx \right)^{\frac{1}{\tilde{\tau}_2}}\nonumber\displaybreak\\
&\leq\left(\int_{\mathbb{R}^N}|G_1(|u|^2)|^{\gamma_1}dx \right)^{\frac{1}{\tilde{\tau}'_1}}\left(\int_{\mathbb{R}^N}|G_1(|u|^2)|^{\gamma_2}dx \right)^{\frac{1}{\tilde{\tau}_1}}\nonumber\\
&\quad+\left(\int_{\mathbb{R}^N}|G_1(|u|^2)|^{\tilde{\gamma}_1}dx \right)^{\frac{1}{\tilde{\tau}'_2}}\left(\int_{\mathbb{R}^N}|G_1(|u|^2)|^{\tilde{\gamma}_2}dx \right)^{\frac{1}{\tilde{\tau}_2}}\nonumber\\
&\leq \sum_{j=1}^2(c_j\int_{\mathbb{R}^N}|u|^2dx)^{\frac{1}{\tilde{\tau}'_j}}\left(c'_j\int_{\mathbb{R}^N}[|u|+h(|u|^2)]^{2^*}dx \right)^{\frac{1}{\tilde{\tau}_j}}\nonumber\\
&\leq \sum_{j=1}^2(c_j\|u_0\|^2_{L^2})^{\frac{1}{\tilde{\tau}'_j}}(2^{2^*-1}c'_jC_s)^{\frac{1}{\tilde{\tau}_j}}\left(\int_{\mathbb{R}^N}[|\nabla u|^2+|\nabla h(|u|^2)|^2dx \right)^{\frac{2^*}{2\tilde{\tau}_j}}.\label{63x1}
\end{align}
Here
\begin{align}
\frac{1}{\tilde{\tau}_1}&=\frac{1-\gamma_1}{\gamma_2-\gamma_1},\quad \frac{1}{\tilde{\tau}'_1}=\frac{\gamma_2-1}{\gamma_2-\gamma_1},\qquad
\frac{1}{\tilde{\tau}_2}=\frac{1-\tilde{\gamma}_1}{\tilde{\gamma}_2-\tilde{\gamma}_1},\quad
\frac{1}{\tilde{\tau}'_2}=\frac{\tilde{\gamma}_2-1}{\tilde{\gamma}_2-\tilde{\gamma}_1}.\label{36x2}
\end{align}

If
$$
\frac{2^*(1-\gamma_1)}{2(\gamma_2-\gamma_1)}=1,\quad \frac{2^*(1-\tilde{\gamma}_1)}{2(\tilde{\gamma}_2-\tilde{\gamma}_1)}<1,\quad (c_1\|u_0\|^2_{L^2})^{\frac{2}{N}}(2^{2^*-1}c'_1C_s)^{\frac{N-2}{N}}<\frac{1}{4},
$$
applying Young inequality to (\ref{63x1}), we obtain
\begin{align}
\int_{\mathbb{R}^N}|G_1(|u|^2)|dx\leq C+\frac{1}{4}\int_{\mathbb{R}^N}[|\nabla u|^2+|\nabla h(|u|^2)|^2]dx.\label{620x1}
\end{align}
Similarly, if
$$
\frac{2^*(1-\gamma_1)}{2(\gamma_2-\gamma_1)}<1,\quad\frac{2^*(1-\tilde{\gamma}_1)}{2(\tilde{\gamma}_2-\tilde{\gamma}_1)}=1,\quad (c_2\|u_0\|^2_{L^2})^{\frac{2}{N}}(2^{2^*-1}c'_2C_s)^{\frac{N-2}{N}}<\frac{1}{4}
$$
applying Young inequality to (\ref{63x1}), we get
\begin{align}
\int_{\mathbb{R}^N}|G(|u|^2)|dx\leq C+\frac{1}{4}\int_{\mathbb{R}^N}[|\nabla u|^2+|\nabla h(|u|^2)|^2]dx.\label{620x2}
\end{align}
If
$$\frac{2^*(1-\gamma_1)}{2(\gamma_2-\gamma_1)}=1,\quad \frac{2^*(1-\tilde{\gamma}_1)}{2(\tilde{\gamma}_2-\tilde{\gamma}_1)}=1,\quad
\sum_{j=1}^2(c_j\|u_0\|^2_{L^2})^{\frac{2}{N}}(2^{2^*-1}c'_jC_s)^{\frac{N-2}{N}}<\frac{1}{4},
$$
(\ref{63x1}) becomes
\begin{align}
\int_{\mathbb{R}^N}|G_1(|u|^2)|dx&\leq \sum_{j=1}^2(c_j\|u_0\|^2_{L^2})^{\frac{1}{\tilde{\tau}'_j}}(c'_jC_s)^{\frac{1}{\tilde{\tau}_j}}\int_{\mathbb{R}^N}[\|\nabla u|^2+|\nabla h(|u|^2)|^2]dx\nonumber\\
&<\frac{1}{4}\int_{\mathbb{R}^N}[|\nabla u|^2+|\nabla h(|u|^2)|^2dx.\label{620x3}
\end{align}
Noticing (\ref{620x1})--(\ref{620x3}), in any case, we have
\begin{align}
&\quad \frac{1}{2}\int_{\mathbb{R}^N}[|\nabla u|^2+\delta_h|\nabla h(|u|^2)|^2+|V(x)||u|^2+|G_2(|u|^2)|]dx+\frac{1}{4}\int_{\mathbb{R}^N}(|W|*|u|^2)|u|^2dx\nonumber\\
&=E(u_0) +\frac{1}{2}\int_{\mathbb{R}^N}|G_1(|u|^2)|dx\leq C+\frac{1}{4}\int_{\mathbb{R}^N}[|\nabla u|^2+\delta_h|\nabla h(|u|^2)|^2]dx,\label{628x1}
\end{align}
which implies that
$$
\int_{\mathbb{R}^N}|\nabla u|^2dx+\int_{\mathbb{R}^N}|\nabla h(|u|^2)|^2dx\leq C.
$$
Theorem 1 is proved. \hfill $\Box$

Now we state pseudo-conformal conservation law as follows.

{\bf Theorem 2.} ({\bf Pseudoconformal conservation law}) {\it Let $u(x,t)$ be the global solution of (\ref{1}) in energy space $X$, $u_0\in X$ and $xu_0\in L^2(\mathbb{R}^N)$. Then
\begin{align}
P(t)&=\int_{\mathbb{R}^N}|(x-2it\nabla)u|^2dx+4t^2\int_{\mathbb{R}^N}\delta_h|\nabla h(|u|^2)|^2dx-4t^2\int_{\mathbb{R}^N}G(|u|^2)dx\nonumber\\
&\quad-4t^2\int_{\mathbb{R}^N}V(x)|u|^2dx-2t^2\int_{\mathbb{R}^N}(W*|u|^2)|u|^2dx\nonumber\\
&=\int_{\mathbb{R}^N}|xu_0|^2dx+4\int_0^t\tau\theta(\tau)d\tau.\label{2211}
\end{align}
Here
\begin{align}
\theta(t)&=\int_{\mathbb{R}^N}-4N\delta_h[2h''(|u|^2)h'(|u|^2)|u|^2+( h'(|u|^2))^2]|u|^2|\nabla u|^2dx\nonumber\\
&\quad-\int_{\mathbb{R}^N} [(N+2)G(|u|^2)-NF(|u|^2)|u|^2]dx\nonumber\\
&\quad-\int_{\mathbb{R}^N}[2V+(x\cdot \nabla V)] |u|^2dx-\int_{\mathbb{R}^N} \left([W+\frac{(x\cdot \nabla W)}{2}]*|u|^2\right)|u|^2dx.\label{2212}
\end{align}
}

{\bf Proof of Theorem 2:} Assume that $u$ is the solution of (\ref{1}), $u_0\in X$ and $xu_0\in L^2(\mathbb{R}^N)$. Using energy conservation law, we get
\begin{align}
P(t)&=\int_{\mathbb{R}^N}|xu|^2dx+4t\Im \int_{\mathbb{R}^N}\bar{u}(x\cdot \nabla u)dx+4t^2\int_{\mathbb{R}^N}|\nabla u|^2dx\nonumber\\
&\quad+4t^2\int_{\mathbb{R}^N}\delta_h|\nabla h(|u|^2)|^2dx-4t^2\int_{\mathbb{R}^N}G(|u|^2)dx\nonumber\\
&\quad-4t^2\int_{\mathbb{R}^N}V(x)|u|^2dx-2t^2\int_{\mathbb{R}^N}(W*|u|^2)|u|^2dx\nonumber\\
&=\int_{\mathbb{R}^N}|xu|^2dx+4t\Im \int_{\mathbb{R}^N}\bar{u}(x\cdot \nabla u)dx+8t^2E(u_0).\label{692}
\end{align}
Recalling that
$$\frac{d}{dt} \int_{\mathbb{R}^N}|x|^2|u|^2dx=-4\Im \int_{\mathbb{R}^N} \bar{u}(x\cdot \nabla u)dx,$$
we get
\begin{align}
P'(t)&=\frac{d}{dt}\int_{\mathbb{R}^N}|xu|^2dx+4\Im \int_{\mathbb{R}^N}\bar{u}(x\cdot \nabla u)dx+4t\frac{d}{dt}\Im \int_{\mathbb{R}^N}\bar{u}(x\cdot \nabla u)dx+16tE(u_0)\nonumber\\
&=4t\frac{d}{dt}\Im \int_{\mathbb{R}^N}\bar{u}(x\cdot \nabla u)dx+16tE(u_0)\nonumber\\
&=4t\left\{-2\int_{\mathbb{R}^N}|\nabla u|^2dx-(N+2)\int_{\mathbb{R}^N}\delta_h|\nabla h(|u|^2)|^2dx\right.\nonumber\\
&\qquad \left.-8N\int_{\mathbb{R}^N}\delta_hh''(|u|^2)h'(|u|^2)|u|^4|\nabla u|^2dx-\int_{\mathbb{R}^N}(x\cdot \nabla V)|u|^2dx\right.\nonumber\\
&\qquad \left.+N\int_{\mathbb{R}^N}[|u|^2F(|u|^2)-G(|u|^2)]dx-\int_{\mathbb{R}^N}[\frac{(x\cdot \nabla W)}{2}*|u|^2]|u|^2dx\right\}\nonumber\\
&\qquad+8t\int_{\mathbb{R}^N}[|\nabla u|^2+\delta_h|\nabla h(|u|^2)|^2-V(x)|u|^2-G(|u|^2)-\frac{1}{2}(W*|u|^2)|u|^2]dx\nonumber\\
&=4t\int_{\mathbb{R}^N}-4N\delta_h[2h''(|u|^2)h'(|u|^2)|u|^2+(h'(|u|^2))^2]|u|^2|\nabla u|^2dx\nonumber\\
&\quad+ 4t\int_{\mathbb{R}^N}[N|u|^2F(|u|^2)-(N+2)G(|u|^2)]dx-4t\int_{\mathbb{R}^N}[2V+(x\cdot \nabla V)] |u|^2dx\nonumber\\
&\quad-4t\int_{\mathbb{R}^N} \left([W+\frac{(x\cdot \nabla W)}{2}]*|u|^2\right)|u|^2dx.\label{693}
\end{align}
Integrating (\ref{693}) from $0$ to $t$, we obtain
$$
P(t)=P(0)+4\int_0^t\tau\theta(\tau)d\tau=\int_{\mathbb{R}^N}|xu_0|^2dx+4\int_0^t\tau\theta(\tau)d\tau.
$$
That is,
\begin{align*}
P(t)&=\int_{\mathbb{R}^N}|(x-2it\nabla) u|^2dx+4t^2\int_{\mathbb{R}^N}\delta_h|\nabla h(|u|^2)|^2dx-4t^2\int_{\mathbb{R}^N}G(|u|^2)dx\nonumber\\
&\quad-4t^2\int_{\mathbb{R}^N}V(x)|u|^2dx-2t^2\int_{\mathbb{R}^N}(W*|u|^2)|u|^2dx\nonumber\\
&=\int_{\mathbb{R}^N}|xu_0|^2dx+4\int_0^t\tau\theta(\tau)d\tau,
\end{align*}
where $\theta(\tau)$ is defined by (\ref{2212}).\hfill$\Box$

\section{Morawetz type estimates based on pseudoconformal conservation law}
\qquad In this section, we will establish Morawetz estimates based on pseudoconformal conservation law.

{\bf Theorem 3.} ({\bf Morawetz type estimates based on pseudoconformal conservation law}) {\it Let $u(x,t)$ be the global solution of (\ref{1})
in energy space $X$, $u_0\in X$ and $xu_0\in L^2(\mathbb{R}^N)$, the space dimension $N\geq 1$ in defocusing case, $N\geq 3$ in combined defocusing and focusing case, $0<M(u_0)<+\infty$ and $0\leq E(u_0)<+\infty$. In addition, suppose that $h(s)\geq 0, \not \equiv 0$, $F(s)=F_1(s)-F_2(s)$ in the combined defocusing and focusing case, $F_1(s)\geq 0$ and $F_2(s)\geq 0$ for $s\geq 0$, and there exist $c_3, c'_3, c_4, c'_4>0$, $0<\gamma_3, \tilde{\gamma}_3<1$ and $\gamma_4,\tilde{\gamma}_4>1$ such that
\begin{align}
&\frac{2^*(1-\gamma_3)}{2(\gamma_4-\gamma_3)}=1,\quad \frac{2^*(1-\tilde{\gamma}_3)}{2(\tilde{\gamma}_4-\tilde{\gamma}_3)}= 1\label{6261'},\\
&C_r(u_0):=\sum_{j=3}^4(c_j\|u_0\|^2_{L^2})^{\frac{2}{N}}(2^{2^*-1}c'_jC_s)^{\frac{N-2}{N}}<1\label{9101}\\
& [|G_1(s)|]^{\gamma_3}\leq c_3s ,\quad [|G_1(s)|]^{\gamma_4}\leq c'_3[h(s)]^{2^*}\ {\rm for}\ 0\leq s\leq 1,\label{6262'}\\
& [|G(s)|]^{\tilde{\gamma}_3}\leq c_4s ,\quad [|G_2(s)|]^{\tilde{\gamma}_4}\leq c'_4[h(s)]^{2^*}\ {\rm for}\  s>1.\label{6263'}
\end{align}

{\bf 1.} Assume that $\delta_h[2h''(s)h'(s)s+(h'(s))^2]\geq 0$, $[(N+2)G_1(s)-NF_1(s)s]\geq 0$ and $[NF_2(s)s-(N+2)G_2(s)]\geq 0$ for $s\geq 0$, $[2V+(x\cdot \nabla V)]\geq 0$ and $[2W+(x\cdot \nabla W)]\geq 0$ for $x\in \mathbb{R}^N$. Then

{\bf Estimate (A):}

\begin{align}
&\int_0^{\infty}\int_{\mathbb{R}^N}\frac{\left[\delta_h|\nabla h(|u|^2)|^2+|V(x)||u|^2+|G_1(|u|^2)|+|G_2(|u|^2)|+\frac{1}{2}(|W|*|u|^2)|u|^2\right]^{\theta}}{a_1(x,t)}dxdt\nonumber\\
&\leq M_1(u_0,\theta),\label{2211'}
\end{align}
where $a_1(x,t)\geq a(x)\geq 0$ for $x\in \mathbb{R}^N$ and $t\geq 0$, $\frac{1}{a(x)}\in L^{\frac{1}{1-\theta}}(\mathbb{R}^N)$, $\frac{1}{2}<\theta<1$.

{\bf Estimate (B):}
\begin{align}
&\quad\int_0^{\infty}\int_{\mathbb{R}^N}\frac{t^2\left[\delta_h|\nabla h(|u|^2)|^2+|V(x)||u|^2+|G_1(|u|^2)|+|G_2(|u|^2)|+\frac{1}{2}(|W|*|u|^2)|u|^2\right]}{a_2(x,t)}dxdt\nonumber\\
&\leq M_2(u_0,k),\label{2222}
\end{align}
where $a_2(x,t)\geq b(x)+t^k$ for $x\in \mathbb{R}^N$ and $t\geq 0$, $1<k<3$ if $b(x)\geq 0$, or $1<k$ if $b(x)\geq b>0$;

Especially, let $b(x)\equiv 0$, $k=2$, then

{\bf Estimate (C): }
\begin{align}
&\quad\int_0^{\infty}\int_{\mathbb{R}^N}\left[\delta_h|\nabla h(|u|^2)|^2+|V(x)||u|^2+|G_1(|u|^2)|+|G_2(|u|^2)|+\frac{1}{2}(|W|*|u|^2)|u|^2\right]dxdt\nonumber\\
&\leq M_3(u_0).\label{2223}
\end{align}

{\bf 2.} Assume that

(i) $-k_1\delta_h(h'(s))^2\leq \delta_h[2h''(s)h'(s)s+(h'(s))^2]\leq 0$ for some $k_1>0$;

(ii) $-k_2|G_1(s)|\leq (N+2)G_1(s)-NF_1(s)s\leq 0$ for some $k_2>0$;

(iii) $-k_3|G_2(s)|\leq NF_2(s)s-(N+2)G_2(s)\leq 0$ for some $k_3>0$;

(iv) $-k_4|V|\leq 2V+(x\cdot \nabla V)\leq 0$ for some $k_4>0$;

(v) $-k_5|W|\leq 2W+(x\cdot \nabla W)\leq 0$ for some $k_5>0$.

Let
\begin{align}
l=\max(Nk_1, k_2, k_3, k_4, k_5).\label{311xj2}
\end{align}

Then

{\bf Estimate (D):}
\begin{align}
&\quad\int_0^{\infty}\int_{\mathbb{R}^N}\frac{t^2\left[\delta_h|\nabla h(|u|^2)|^2+|G_1(|u|^2)|+|G_2(|u|^2)|+|V(x)||u|^2+\frac{1}{2}(|W|*|u|^2)|u|^2\right]}{a_3(x,t)}dxdt\nonumber\\
&\leq M_4(u_0,k,l)\label{2224}
\end{align}
 Here $a_3(x,t)\geq (c(x)+t)^k$ for $x\in \mathbb{R}^N$ and $t\geq 0$, $l+1<k<3$ if $l<2$ in defocusing case, $k>1+\frac{l[1+C_r(u_0)]}{1-C_r(u_0)}$ in combined defocusing and focusing case, if $c(x)\geq 0$.  While $l+1<k$ in defocusing case, $k>1+\frac{l[1+C_r(u_0)]}{1-C_r(u_0)}$ in combinied defocusing and focusing case, if $c(x)\geq c>0$.

Especially, if $c(x)\equiv 0$, $l<1$ and $k=2$, then

{\bf Estimate (E): }
\begin{align}
&\quad\int_0^{\infty}\int_{\mathbb{R}^N}\left[\delta_h|\nabla h(|u|^2)|^2+|G_1(|u|^2)|+|G_2(|u|^2)|+|V(x)||u|^2+\frac{1}{2}(|W|*|u|^2)|u|^2\right]dxdt\nonumber\\
&\leq M_5(u_0,l).\label{2225}
\end{align}
}

We divide this section into two subsection according to Case 1 and Case 2.
\subsection{The proof of Theorem 3 in Case 1}
\qquad In this subsection, we prove Theorem 3 in Case 1.

{\bf The proof of Theorem 3 in Case 1:} First, we give estimates for
\begin{align}
\int_{\mathbb{R}^N}\Phi(V,u,W)dx&:=\int_{\mathbb{R}^N}[\delta_h|\nabla h(|u|^2)|^2+|G_1(|u|^2)|+|G_2(|u|^2)|]dx\nonumber\\
&\quad+\int_{\mathbb{R}^N}[|V(x)||u|^2+\frac{1}{2}(|W|*|u|^2)|u|^2]dx\label{311xj1}
\end{align}
in two subcases.

{\bf Subcase (1)}. Defocusing case, $N\geq 1$. By energy conservation law, we get
\begin{align}
\int_{\mathbb{R}^N}\Phi(V,u,W)dx\leq 2E(u_0)\ {\rm for \ any} \ t\geq 0({\rm especially \ for}\ 0\leq t\leq 1).\label{36xj1}
\end{align}

Using (\ref{2211}) and (\ref{2212}), we have
\begin{align}
4t^2\int_{\mathbb{R}^N}\Phi(V,u,W)dx\leq \int_{\mathbb{R}^N}|xu_0|^2dx, \ \int_{\mathbb{R}^N}\Phi(V,u,W)dx\leq \frac{C(u_0)}{4t^2}\quad
{\rm for } \ t\geq 1.\label{222x2}
\end{align}
Here
\begin{align}
C(u_0)=\int_{\mathbb{R}^N}|xu_0|^2dx.\label{224x1}
\end{align}

{\bf Subcase (2)}. Combined defocusing and focusing case, $N\geq 3$.

By energy conservation law, we get
\begin{align*}
&\quad [1-C_r(u_0)]\int_{\mathbb{R}^N}[|\nabla h(|u|^2)|^2+|V(x)||u|^2+|G_2(|u|^2)|+\frac{1}{2}(|W|*|u|^2)|u|^2]dx\nonumber\\
&\leq\int_{\mathbb{R}^N}[|\nabla h(|u|^2)|^2-|G_1(|u|^2)|+|G_2(|u|^2)+|V(x)||u|^2+\frac{1}{2}(|W|*|u|^2)|u|^2]dx\nonumber\\
&=2E(u_0),
\end{align*}
and
\begin{align}
\int_{\mathbb{R}^N}[|\nabla h(|u|^2)|^2+|V(x)||u|^2+|G_2(|u|^2)|+\frac{1}{2}(|W|*|u|^2)|u|^2]dx\leq\frac{2E(u_0)}{[1-C_r(u_0)]},\label{32xj1}
\end{align}
consequently,
\begin{align}
\int_{\mathbb{R}^N}\Phi(V,u,W)dx\leq \frac{2E(u_0)[1+C_r(u_0)]}{[1-C_r(u_0)]}\quad {\rm for\ any }\ t\geq 0\ ( {\rm  especially\ for} \ 0\leq t\leq 1).\label{321}
\end{align}
Using (\ref{2211}) and (\ref{2212}), we obtain
\begin{align*}
&\quad[1-C_r(u_0)] \left(4t^2\int_{\mathbb{R}^N}[|\nabla h(|u|^2)|^2+|V(x)||u|^2+|G_2(|u|^2)|]dx+2t^2\int_{\mathbb{R}^N}(|W|*|u|^2)|u|^2dx\right)\nonumber\\
&\leq[1-C_r(u_0)] 4t^2\int_{\mathbb{R}^N}|\nabla h(|u|^2)|^2dx+4t^2\int_{\mathbb{R}^N}[|V(x)||u|^2+|G_2(|u|^2)|]dx\nonumber\\
&\quad+2t^2\int_{\mathbb{R}^N}(|W|*|u|^2)|u|^2dx\nonumber\\
&\leq \int_{\mathbb{R}^N}|xu_0|^2dx,
\end{align*}
which implies that
\begin{align}
&\quad\int_{\mathbb{R}^N}[|\nabla h(|u|^2)|^2+|V(x)||u|^2+|G_2(|u|^2)|+\frac{1}{2}(|W|*|u|^2)|u|^2]dx\nonumber\\
&\leq \frac{C(u_0)}{4[1-C_r(u_0)]t^2},\label{36xj2}
\end{align}
and consequently
\begin{align}
\int_{\mathbb{R}^N}\Phi(V,u,W)dx\leq \frac{C(u_0)[1+C_r(u_0)]}{4[1-C_r(u_)]t^2} \quad {\rm for \ any} \quad t\geq 1.\label{222x2'}
\end{align}

Now Morawetz estimates can be proved below.

{\bf Estimate (A):}

For any $\frac{1}{2}<\theta<1$, $a_1(x,t)\geq a(x)$ and $\frac{1}{a(x)}\in L^{\frac{1}{1-\theta}}(\mathbb{R}^N)$, using (\ref{36xj1})--(\ref{222x2'}), we have
\begin{align}
\int_0^{\infty}\int_{\mathbb{R}^N}\frac{\left[\Phi(V,u,W)\right]^{\theta}}{a_1(x,t)}dxdt&\leq  \int_0^{\infty}\int_{\mathbb{R}^N}\frac{\left[\Phi(V,u,W)\right]^{\theta}}{a(x)}dxdt\nonumber\\
&=\int_0^1\int_{\mathbb{R}^N}\frac{\left[\Phi(V,u,W)\right]^{\theta}}{a(x)}dxdt
+\int_1^{\infty}\int_{\mathbb{R}^N}\frac{\left[\Phi(V,u,W)\right]^{\theta}}{a(x)}dxdt\nonumber\\
&\leq\int_0^1\left(\int_{\mathbb{R}^N}\Phi(V,u,W)dx\right)^{\theta}\left(\int_{\mathbb{R}^N}\frac{1}{[a(x)]^{\frac{1}{1-\theta}}}dx\right)^{1-\theta}dt\nonumber\\
&\quad +\int_1^{\infty}\left(\int_{\mathbb{R}^N}\Phi(V,u,W)dx\right)^{\theta}\left(\int_{\mathbb{R}^N}\frac{1}{[a(x)]^{\frac{1}{1-\theta}}}dx\right)^{1-\theta}dt\nonumber\\
&\leq \left[\int_0^1 Cdt+\int_1^{\infty}\frac{C'}{t^{2\theta}}dt\right]\left(\int_{\mathbb{R}^N}\frac{1}{[a(x)]^{\frac{1}{1-\theta}}}dx\right)^{1-\theta}\nonumber\\
&\leq M_1(u_0,\theta).\label{222w1}
\end{align}
Here
\begin{equation*}
M_1(u_0,\theta)=\left\{
\begin{array}{lll}
\left([2E(u_0)]^{\theta}
+\frac{1}{2\theta-1}[\frac{C(u_0)}{4}]^{\theta}\right) \left(\int_{\mathbb{R}^N}\frac{1}{[a(x)]^{\frac{1}{1-\theta}}}dx\right)^{1-\theta}\ {\rm in\ defocusing \ case}\\
\left(\frac{1+C_r(u_0)}{1-C_r(u_0)}\right)^{\theta}\left([2E(u_0)]^{\theta}
+\frac{1}{2\theta-1}[\frac{C(u_0)}{4}]^{\theta}\right) \left(\int_{\mathbb{R}^N}\frac{1}{[a(x)]^{\frac{1}{1-\theta}}}dx\right)^{1-\theta}\quad {\rm in}\\
\quad {\rm \ combined\ defocusing\ and\ focusing\ case}.
\end{array}\right.
\end{equation*}

{\bf Estimate (B):}

If $b(x)\geq 0$, $1<k<3$, we get
\begin{align}
\int_0^{\infty}\int_{\mathbb{R}^N}\frac{t^2\Phi(V,u,W)}{a_2(x,t)}dxdt&\leq\int_0^{\infty}\int_{\mathbb{R}^N}\frac{t^2\Phi(V,u,W)}{b(x)+t^k}dxdt\nonumber\\
&=\int_0^1\int_{\mathbb{R}^N}\frac{t^2\Phi(V,u,W)}{b(x)+t^k}dxdt
+\int_1^{\infty}\int_{\mathbb{R}^N}\frac{t^2\Phi(V,u,W)}{b(x)+t^k}dxdt\nonumber\\
&\leq \int_0^1t^{2-k}\int_{\mathbb{R}^N}\Phi(V,u,W)dxdt+\int_1^{\infty}\frac{1}{t^k}\int_{\mathbb{R}^N}t^2\Phi(V,u,W)dxdt\nonumber\\
&\leq \int_0^1 Ct^{2-k}dt+\int_1^{\infty}\frac{C'}{t^k}dt\leq M_2(u_0,k).
\end{align}

If $b(x)\geq b>0$, $1<k$, we obtain
\begin{align}
\int_0^{\infty}\int_{\mathbb{R}^N}\frac{t^2\Phi(V,u,W)}{a_2(x,t)}dxdt&\leq \int_0^{\infty}\int_{\mathbb{R}^N}\frac{t^2\Phi(V,u,W)}{b(x)+t^k}dxdt\nonumber\\
&\leq \int_0^1\frac{Ct^2}{b}dt+\int_1^{\infty}\frac{C'}{t^k}dt\leq M_2(u_0,k).\label{222w2}
\end{align}
 Here
\begin{equation}
\label{M(2)}M_2(u_0,k)=\left\{
\begin{array}{lll}
\frac{2E(u_0)}{3-k}+\frac{C(u_0)}{4(k-1)}\quad {\rm in\ defocusing \ case}\\
\frac{1+C_r(u_0)}{1-C_r(u_0)}[\frac{2E(u_0)}{3-k}+\frac{C(u_0)}{4(k-1)}]\quad {\rm in\ combined}\\
\quad {\rm defocusing\ and\ focusing\ case}
\end{array}\right.
\end{equation}
if $b(x)\geq 0$, $1<k<3$. While if $b(x)\geq b>0$, $1<k$,
\begin{equation}
\label{M(2')}M_2(u_0,k)=\left\{
\begin{array}{lll}
\frac{2E(u_0)}{3b}+\frac{C(u_0)}{4(k-1)}\quad {\rm in\ defocusing \ case}\\
\frac{1+C_r(u_0)}{1-C_r(u_0)}[\frac{2E(u_0)}{3b}+\frac{C(u_0)}{4(k-1)}]\quad {\rm in\ combined}\\
\quad {\rm  defocusing\ and\ focusing\ case}.
\end{array}\right.
\end{equation}

 Especially, if $b(x)\equiv0$ and $k=2$, we have

 {\bf Estimate (C):}
\begin{align}
&\quad\int_0^{\infty}\int_{\mathbb{R}^N}\left[\delta_h|\nabla h(|u|^2)|^2+|G(|u|^2)|+|V(x)||u|^2+\frac{1}{2}(|W|*|u|^2)|u|^2\right]dxdt\nonumber\\
&\leq M_3(u_0).\label{222w3}
\end{align}
Here
\begin{equation}
M_3(u_0)=\left\{
\begin{array}{lll}
2E(u_0)+\frac{C(u_0)}{4}\quad {\rm in\ defocusing \ case}\\
\frac{1+C(u_0)}{1-C_r(u_0)}[2E(u_0)+\frac{C(u_0)}{4}]\quad {\rm in\ combined}\\
\quad {\rm  defocusing\ and\ focusing\ case}
\end{array}\right.
\end{equation}

\subsection{The proof of Theorem 3 in Case 2}
\qquad In this subsection, we prove Theorem 3 in Case 2.

{\bf The proof of Theorem 3 in Case 2:}

{\bf Estimate (D):} We prove it in two subcases.

{\bf Subcase (i).} Defocusing case, $N\geq 1$. By energy conservation law, we also have
$$
\int_{\mathbb{R}^N}\Phi(V,u,W)dx\leq 2E(u_0)\quad  {\rm for \ any} \ t\geq 0({\rm especially\ for}\ 0\leq t\leq 1).
$$

Letting
\begin{align}
A(t)=&4\int_0^t\tau \int_{\mathbb{R}^N}\Phi(V,u,W)dxd\tau,\label{222x3}
\end{align}
 using (\ref{2211}) and (\ref{2212}), we have
\begin{align*}
tA'(t)\leq \int_{\mathbb{R}^N}|xu_0|^2dx+l A(t)=C(u_0)+l A(t),
\end{align*}
i.e.,
\begin{align}
A'(t)\leq \frac{l}{t} A(t)+\frac{C(u_0)}{t}.\label{222x4}
\end{align}
Applying Gronwall inequality to (\ref{222x4}), we get
\begin{align}
A(t)&\leq e^{\int_1^t\frac{l}{\eta}d\eta}[A(1)+\int_1^t\frac{C(u_0)}{\eta}e^{-\int_1^{\eta}\frac{l}{\xi}d\xi}d\eta]\nonumber\\
&=t^l[A(1)+\frac{C(u_0)}{l}-\frac{C(u_0)}{lt^l}]\leq [4E(u_0)+\frac{C(u_0)}{l}]t^l\label{2241}
\end{align}
for any $t\geq1$. (\ref{222x4}) and (\ref{2241}) mean that
\begin{align}
\int_{\mathbb{R}^N}\Phi(V,u,W)dx&\leq \frac{C(u_0)}{4t^2}+ \frac{[4lE(u_0)+C(u_0)]}{4t^{2-l}}\quad {\rm for\ any} \ t\geq 1.\label{222x5}
\end{align}

In defocusing case, we obtain
\begin{align}
&\quad \int_0^{\infty}\int_{\mathbb{R}^N}\frac{t^2\Phi(V,u,W)}{a_3(x,t)}dxdt\nonumber\\
&\leq \int_0^{\infty}\int_{\mathbb{R}^N}\frac{t^2\Phi(V,u,W)}{(c(x)+t)^k}dxdt\nonumber\\
&=\int_0^1\int_{\mathbb{R}^N}\frac{t^2\Phi(V,u,W)}{(c(x)+t)^k}dxdt +\int_1^{\infty}\int_{\mathbb{R}^N}\frac{t^2\Phi(V,u,W)}{(c(x)+t)^k}dxdt\nonumber\\
&\leq \int_0^1t^{2-k}\int_{\mathbb{R}^N}\Phi(V,u,W)dxdt+\int_1^{\infty}\frac{1}{t^k}\int_{\mathbb{R}^N}t^2\Phi(V,u,W)dxdt\nonumber\\
&\leq 2E(u_0)\int_0^1t^{2-k}dt+\frac{C(u_0)}{4}\int_1^{+\infty}\frac{1}{t^k}dt\nonumber\\
&\quad+\frac{[4lE(u_0)+C(u_0)]}{4}\int_1^{\infty}\frac{1}{t^{k-l}}dt\nonumber\\
&\leq \frac{2E(u_0)}{3-k}+\frac{C(u_0)}{4(k-1)}+\frac{4lE(u_0)+C(u_0)}{4[k-(l+1)]}\label{222w2}
\end{align}
for $c(x)\geq 0$, $l+1<k<3$ if $l<2$.

Similarly, we have
\begin{align}
\int_0^{\infty}\int_{\mathbb{R}^N}\frac{t^2\Phi(V,u,W)}{a_3(x,t)}dxdt&\leq \int_0^{\infty}\int_{\mathbb{R}^N}\frac{t^2\Phi(V,u,W)}{(c(x)+t)^k}dxdt\nonumber\\
&\leq 2E(u_0)\int_0^1\frac{t^2}{(c+t)^k}dt+\frac{C(u_0)}{4}\int_1^{+\infty}\frac{1}{t^k}dt\nonumber\\
&\quad+\frac{[4lE(u_0)+C(u_0)]}{4}\int_1^{\infty}\frac{1}{t^{k-l}}dt\nonumber\\
&\leq \frac{2E(u_0)}{3c^k}+\frac{C(u_0)}{4(k-1)}+\frac{4lE(u_0)+C(u_0)}{4[k-(l+1)]}\label{222w2'}
\end{align}
for $c(x)\geq c>0$, $l+1<k$.

{\bf Subcase (ii).} Combined defocusing and focusing case, $N\geq 3$. Recall that (\ref{321})
\begin{align*}
\int_{\mathbb{R}^N}\Phi(V,u,W)dx\leq \frac{2E(u_0)[1+C_r(u_0)]}{[1-C_r(u_0)]}
\end{align*}
for any $t\geq 0$(especially for $0<t\leq 1$).

 Using (\ref{2211}) and (\ref{2212}), we get
\begin{align}
&\quad[1-C_r(u_0)] 4t^2\int_{\mathbb{R}^N}|\nabla h(|u|^2)|^2dx+4t^2\int_{\mathbb{R}^N}[|V(x)||u|^2+|G_2(|u|^2)|]dx\nonumber\\
&\qquad+2t^2\int_{\mathbb{R}^N}(|W|*|u|^2)|u|^2dx\nonumber\\
&\leq C(u_0)+4l\int_0^t\tau \int_{\mathbb{R}^N}\Phi(V,u,W)dxd\tau\nonumber\\
&\leq C(u_0)+4l[1+C_r(u_0)] \int_0^t\tau\int_{\mathbb{R}^N}[|\nabla h(|u|^2)|^2dx+|V(x)||u|^2+|G_2(|u|^2)|\nonumber\\
&\qquad+\frac{1}{2}(|W|*|u|^2)|u|^2]dxd\tau.\label{32x1}
\end{align}
Letting
\begin{align*}
B(t)=4\int_0^t\tau\int_{\mathbb{R}^N}[|\nabla h(|u|^2)|^2dx+|V(x)||u|^2+|G_2(|u|^2)|+\frac{1}{2}(|W|*|u|^2)|u|^2]dxd\tau,
\end{align*}
we have from (\ref{32x1})
\begin{align}
B'(t)\leq \frac{C(u_0)}{[1-C_r(u_0)]t}+\frac{l[1+C_r(u_0)]}{[1-C_r(u_0)]t}B(t).\label{32x2}
\end{align}
Applying Gronwall inequality to (\ref{32x2}), and using (\ref{32xj1}), we obtain
\begin{align*}
B(t)\leq [\frac{4lE(u_0)[1+C_r(u_0)]+C(u_0)[1-C_r(u_0)]}{l[1-C^2_r(u_0)]}]t^{\frac{l[1+C_r(u_0)]}{1-C_r(u_0)}},
\end{align*}
and
\begin{align}
&\quad\int_{\mathbb{R}^N}|\nabla h(|u|^2)|^2+|V(x)||u|^2+\frac{1}{2}(|W|*|u|^2)|u|^2]dx\nonumber\\
&\leq \frac{C(u_0)}{4[1-C_r(u_0)]t^2}+\frac{4lE(u_0)[1+C_r(u_0)]+C(u_0)[1-C_r(u_0)]}{4[1-C_r(u_0)]^2t^{2-\frac{l[1+C_r(u_0)]}{1-C_r(u_0)}}}\label{36xj3}
\end{align}
for $t\geq 1$. Consequently,
\begin{align}
\int_{\mathbb{R}^N}\Phi(V,u,W)dx\leq \frac{[1+C_r(u_0)]}{4[1-C_r(u_0)]}\left(\frac{C(u_0)}{t^2}+\frac{4lE(u_0)[1+C_r(u_0)]
+C(u_0)[1-C_r(u_0)]}{[1-C_r(u_0)]t^{2-\frac{l[1+C_r(u_0)]}{1-C_r(u_0)}}}\right) \label{222x2''}
\end{align}
for any $t\geq 1$.

Similar to (\ref{222w2}), in combined defocusing and focusing case,
\begin{align}
&\quad\int_0^{\infty}\int_{\mathbb{R}^N}\frac{t^2\Phi(V,u,W)}{a_3(x,t)}dxdt\leq \int_0^{\infty}\int_{\mathbb{R}^N}\frac{t^2\Phi(V,u,W)}{(c(x)+t)^k}dxdt\nonumber\\
&\leq \frac{2E(u_0)[1+C_r(u_0)]}{[1-C_r(u_0)]}\int_0^1t^{2-k}dt\nonumber\\
&\quad+\int_1^{\infty}\frac{C(u_0)[1+C_r(u_0)]}{4[1-C_r(u_0)]t^{k}}+
\frac{4lE(u_0)[1+C_r(u_0)]^2+C(u_0)[1-C^2_r(u_0)]}{4[1-C_r(u_0)]^2}\frac{1}{t^{k-\frac{l[1+C_r(u_0)]}{1-C_r(u_0)}}}dt\nonumber\\
&=\frac{[1+C_r(u_0)]}{[1-C_r(u_0)]}\left(\frac{2E(u_0)}{(3-k)}+\frac{C(u_0)}{4(k-1)}
+\frac{4lE(u_0)[1+C_r(u_0)]+C(u_0)[1-C_r(u_0)]}{4\{(k-1)[1-C_r(u_0)]-l[1+C_r(u_0)]\}}\right)\label{32x4}
\end{align}
for $c(x)\geq 0$, $k>1+\frac{l[1+C_r(u_0)]}{1-C_r(u_0)}$. Combining (\ref{222w2}) and (\ref{32x4}), we have
\begin{align}
\int_0^{\infty}\int_{\mathbb{R}^N}\frac{t^2\Phi(V,u,W)}{a_3(x,t)}dxdt\leq \int_0^{\infty}\int_{\mathbb{R}^N}\frac{t^2\Phi(V,u,W)}{(c(x)+t)^k}dxdt\leq M_4(u_0,k,l).
\end{align}
 Here
\begin{equation}
\label{M(4)}M_4(u_0,k,l)=\left\{
\begin{array}{lll}
\frac{2E(u_0)}{3-k}+\frac{C(u_0)}{4(k-1)}+\frac{4lE(u_0)+C(u_0)}{4l[k-(l+1)]}\quad {\rm in\ defocusing \ case}\\
\frac{[1+C_r(u_0)]}{[1-C_r(u_0)]}\left(\frac{2E(u_0)}{(3-k)}+\frac{C(u_0)}{4(k-1)}
+\frac{4lE(u_0)[1+C_r(u_0)]+C(u_0)[1-C_r(u_0)]}{4\{(k-1)[1-C_r(u_0)]-l[1+C_r(u_0)]\}}\right)\\
 \quad {\rm in\ combined}\ {\rm defocusing\ and\ focusing\ case}
\end{array}\right.
\end{equation}
if $c(x)\geq 0$.

Similarly to (\ref{222w2'}), (\ref{32x4}) in combined defocusing and focusing case,
\begin{align}
&\quad \int_0^{\infty}\int_{\mathbb{R}^N}\frac{t^2\Phi(V,u,W)}{a_3(x,t)}dxdt\leq \int_0^{\infty}\int_{\mathbb{R}^N}\frac{t^2\Phi(V,u,W)}{(c(x)+t)^k}dxdt\nonumber\\
 &\leq \frac{[1+C_r(u_0)]}{[1-C_r(u_0)]}\left(\frac{2E(u_0)}{3c^k}+\frac{C(u_0)}{4(k-1)}+\frac{4lE(u_0)[1+C_r(u_0)]+C(u_0)[1-C_r(u_0)]}{4\{(k-1)[1-C_r(u_0)]-l[1+C_r(u_0)]\}}\right)\label{32xxj2}
\end{align}
for $c(x)\geq c>0$, $k>1+\frac{l[1+C_r(u_0)]}{1-C_r(u_0)}$. Combining (\ref{222w2'}) and (\ref{32xxj2}), we get
\begin{align}
\int_0^{\infty}\int_{\mathbb{R}^N}\frac{t^2\Phi(V,u,W)}{a_3(x,t)}dxdt\leq \int_0^{\infty}\int_{\mathbb{R}^N}\frac{t^2\Phi(V,u,W)}{(c(x)+t)^k}dxdt\leq M_4(u_0,k,l).
\end{align}
Here
\begin{equation}
\label{M(4')}M_4(u_0,k,l)=\left\{
\begin{array}{lll}
\frac{2E(u_0)}{3c^k}+\frac{4lE(u_0)+C(u_0)}{4l[k-(l+1)]}\quad {\rm in\ defocusing \ case}\\
\frac{[1+C_r(u_0)]}{[1-C_r(u_0)]}\left(\frac{2E(u_0)}{3c^k}+\frac{C(u_0)}{4(k-1)}
+\frac{4lE(u_0)[1+C_r(u_0)]+C(u_0)[1-C_r(u_0)]}{4\{(k-1)[1-C_r(u_0)]-l[1+C_r(u_0)]\}}\right)\\
 \quad {\rm in\ combined} \quad {\rm defocusing\ and\ focusing\ case}
\end{array}\right.
\end{equation}
if $c(x)\geq c>0$.

{\bf Estimate (E):}

 Especially, if $c(x)\equiv 0$, $k=2$, $l<\frac{1-C_r(u_0)}{1+C_r(u_0)}$, by the discussions above, we have
\begin{align}
\int_0^{\infty}\int_{\mathbb{R}^N}\Phi(V,u,W)dxdt\leq M_5(u_0,l).\label{222w3}
\end{align}
Here
\begin{equation}
\label{M(5)}M_5(u_0,l)=\left\{
\begin{array}{lll}
2E(u_0)+\frac{C(u_0)}{4}+\frac{4lE(u_0)+C(u_0)}{4l(1-l)}\quad {\rm in\ defocusing \ case}\\
\frac{[1+C_r(u_0)]}{[1-C_r(u_0)]}\left(2E(u_0)+\frac{C(u_0)}{4}+\frac{4lE(u_0)[1+C_r(u_0)]+C(u_0)[1-C_r(u_0)]}{4\{[1-C_r(u_0)]-l[1+C_r(u_0)]\}}\right)\\
 \quad {\rm in\ combined}\ {\rm defocusing\ and\ focusing\ case}
\end{array}\right.
\end{equation}

{\bf Remark 3.1.} 1. The assumptions of Case 2 can be weaken as: Assume that at least one of (i)--(iv) holds. And the corresponding value of $l$ can be take one of $Nk_1$, $k_2$, $k_3$ and $2k_4$. For example, if (i) holds, while $ NF(s)s-(N+2)G(s)\geq 0$, $2V+(x\cdot \nabla V)\geq 0$,
$2W+(x\cdot \nabla W)\geq 0$, we can take $l=Nk_1$; If (i) and (ii) hold, while $2V+(x\cdot \nabla V)\geq 0$, $2W+(x\cdot \nabla W)\geq 0$, we can take $l=\max(Nk_1,k_2)$, and so on.

2. By the proof of Theorem 3, in defocusing case, we obtain
\begin{align}
\int_{\mathbb{R}^N}|(x-2it\nabla)u|^2dx\leq C\quad {\rm in\ Case\ 1},\quad \int_{\mathbb{R}^N}|(x-2it\nabla)u|^2dx\leq Ct^l\quad {\rm in\ Case\ 2}.\label{93xji1}
\end{align}

\section{Spacetime bound estimates based on pseudoconformal conservation law}
\qquad In this section, we will establish spacetime bound estimates based on pseudoconformal conservation law.

{\bf Theorem 4.} ({\bf Space-time bounds based on pseudo-conformal conservation law}) {\it Let  $u(x,t)$ be the solution of (\ref{1})
in energy space $X$, $u_0\in X$ and $xu_0\in L^2(\mathbb{R}^N)$, the space dimension $N\geq 1$ in defocusing case, $N\geq 3$ in combined defocusing and focusing case, $0<M(u_0)<+\infty$ and $0\leq E(u_0)<+\infty$. Assume that $V(x)$, $h(s)$, $F(s)$, $G(s)$ and $W(x)$ satisfy the assumptions of Theorem 3. Then

{\bf Bound (F):} Weighted spacetime bound
\begin{align}
&\left(\int_0^{+\infty}\left(\int_{\mathbb{R}^N}w(x,t)\left[\Phi(V,u,W)\right]^{\theta}dx\right)^pdt\right)^{\frac{1}{p}}\leq C(u_0,p,\theta).\label{37w1}
\end{align}
Here
\begin{align}
\Phi(V,u, W)=\delta_h|\nabla h(|u|^2)|^2+|G_1(|u|^2)|+|G_2(|u|^2)|+|V(x)||u|^2+\frac{1}{2}(|W|*|u|^2)|u|^2.\label{618w1}
\end{align}
 $0<\theta\leq 1$,  $w(x,t)$ satisfies (w1) $0\leq w(x,t)\leq c_w$ for all $x\in\mathbb{R}^N$ and $t\geq 0$ if $\theta=1$, or (w2) $0\leq w(x,t)$ for all $x\in\mathbb{R}^N$ and $t\geq 0$, $\int_{\mathbb{R}^N}|w(x,t)|^{\frac{1}{1-\theta}}dx\leq c'_w$ if $0<\theta<1$, $p>\frac{1}{2\theta}$ in defocusing case, and
$$
p>\max\left(\frac{1}{2\theta}, \frac{[1-C_r(u_0)]}{\theta [2(1-C_r(u_0))-l(1+C_r(u_0))]}\right),\quad 0<l<\frac{2[1-C_r(u_0)]}{[1+C_r(u_0)]}
$$
in combined defocusing and focusing case.

Moreover, if $N\geq 3$, then

{\bf Bound (G):} Weighted spacetime norm
\begin{align}
\|G_1(|u|^2)\|_{L_w^q(\mathbb{R}^+)L_w^r(\mathbb{R}^N)}&=\left(\int_0^{+\infty}\left(\int_{\mathbb{R}^N}w(x,t)[|G_1(|u|^2)|]^rdx\right)^{\frac{q}{r}}dt\right)^{\frac{1}{q}}\nonumber\\
&\leq C(u_0,r,q,\gamma_1,\gamma_2,\tilde{\gamma}_1,\tilde{\gamma}_2).\label{37w2}
\end{align}
Here $1\leq r<\gamma_2$, $1\leq r<\tilde{\gamma}_2$, $w(x,t)$ satisfies (w1) $0\leq w(x,t)\leq c_v$ for all $x\in\mathbb{R}^N$ and $t\geq 0$ if $1\leq r<\gamma_2$, $1\leq r<\tilde{\gamma}_2$, or (w2) $0\leq w(x,t)$ for all $x\in\mathbb{R}^N$ and $t\geq 0$,  $\int_{\mathbb{R}^N}|w(x,t)|^{\frac{\delta}{\delta-1}}dx\leq c'_w$ for some $1<\delta<\frac{\gamma_2}{r}\leq\gamma_2$ and $1<\delta<\frac{\tilde{\gamma}_2}{r}\leq\tilde{\gamma}_2$.
$$q>\frac{r\sigma(\gamma_2-\gamma_1)}{2^*(r\sigma-\gamma_1)},\quad q>\frac{r\sigma(\tilde{\gamma}_2-\tilde{\gamma}_1)}{2^*(r\sigma-\tilde{\gamma}_1)}$$  for combined defocusing and focusing subcase of Case 1 in Theorem 3,
$$q>\frac{2r\sigma(\gamma_2-\gamma_1)[1-C_r(u_0)]}{2^*(r\sigma-\gamma_1)[2(1-C_r(u_0))-l(1+C_r(u_0))]},$$
$$
q>\frac{2r\sigma(\tilde{\gamma}_2-\tilde{\gamma}_1)[1-C_r(u_0)]}{2^*(r\sigma-\tilde{\gamma}_1)[2(1-C_r(u_0))-l(1+C_r(u_0))]},
$$
$0<l<\frac{2[1-C_r(u_0)]}{[1+C_r(u_0)]}$ for combined defocusing and focusing subcase of Case 2 in Theorem 3, where $\sigma=1$ if (w1) holds, while $\sigma=\delta$ if (w2) holds.
}

{\bf Proof of Theorem 4:} Similar to (\ref{63x1}),  we get
\begin{align}
\int_{\mathbb{R}^N}|G_1(|u|^2)|dx&\leq \sum_{j=1}^2(c_j\|u_0\|^2_{L^2})^{\frac{1}{\tilde{\tau}'_j}}(2^{2^*-1}c'_jC_s)^{\frac{1}{\tilde{\tau}_j}}\int_{\mathbb{R}^N}|\nabla h(|u|^2)|^2dx\nonumber\\
&:=C_r(u_0)\int_{\mathbb{R}^N}|\nabla h(|u|^2)|^2dx.\label{371}
\end{align}
if $N\geq 3$. $\tilde{\tau}_1$, $\tilde{\tau}'_1$, $\tilde{\tau}_2$ and $\tilde{\tau}'_2$ are the same as those in (\ref{36x2}).

{\bf Bound (F):} We will prove (\ref{37w1}) in three cases. We only give the details in Case (I), the proofs in Case (II) and Case(II)
are similar to that in Case (I).

{\bf Case (I).} Defocusing subcase in Case 2 of Theorem 3. In this case,
\begin{align*}
&\int_{\mathbb{R}^N}\Phi(V,u,W)dx\leq 2E(u_0)\quad {\rm for}\quad 0\leq t\leq 1,\\
&\int_{\mathbb{R}^N}\Phi(V,u,W)dx\leq \frac{C(u_0)}{4t^2}+\frac{[4lE(u_0)+C(u_0)]}{4t^{2-l}}\quad {\rm for}\quad t>1.
\end{align*}
We discuss it in two subcases.

{\bf Subcase (i).} $0\leq w(x,t)\leq c_w$ for all $x\in\mathbb{R}^N$ and $t\geq 0$ if $\theta=1$. By (\ref{36xj1}) and (\ref{222x5}), we obtain
\begin{align}
&\quad\left(\int_0^{+\infty}\left(\int_{\mathbb{R}^N}w(x,t)\Phi(V,u,W)dx\right)^pdt\right)^{\frac{1}{p}}\nonumber\\
&\leq c_w\left(\int_0^1[2E(u_0)]^pdt+\int_1^{+\infty}\left(\frac{C(u_0)}{4t^2}+\frac{4lE(u_0)+C(u_0)}{4t^{(2-l)}}\right)^pdt\right)^{\frac{1}{p}}\nonumber\\
&\leq c_w\tilde{c}_1\left(\int_0^1[2E(u_0)]^pdt\right)^{\frac{1}{p}}+\tilde{c}_1\left(\int_1^{+\infty}\left(\frac{C(u_0)}{4t^2}+\frac{4lE(u_0)+C(u_0)}{4t^{(2-l)}}\right)^pdt\right)^{\frac{1}{p}}\nonumber\\
&\leq 2E(u_0)\tilde{c}_1c_w+\frac{c_w\tilde{c}^2_1\tilde{c}_2C(u_0)}{4}\left(\int_1^{+\infty}\frac{1}{t^{2p}}dt\right)^{\frac{1}{p}}+
\frac{c_w\tilde{c}^2_1\tilde{c}_2[4lE(u_0)+C(u_0)]}{4}\left(\int_1^{+\infty}\frac{1}{t^{(2-l)p}}dt\right)^{\frac{1}{p}}\nonumber\\
&\leq 2E(u_0)\tilde{c}_1c_w+\frac{c_w\tilde{c}^2_1\tilde{c}_2C(u_0)}{4[2p-1]^{\frac{1}{p}}}
+\frac{c_w\tilde{c}^2_1\tilde{c}_2[4lE(u_0)+C(u_0)]}{4[(2-l)p-1]^{\frac{1}{p}}}:=C_1(u_0,p).\label{38x3}
\end{align}
Here $\tilde{c}_1, \tilde{c}_2=1$ if $p<1$, $\tilde{c}_1, \tilde{c}_2=2^{\frac{p-1}{p}}$ if $p\geq 1$.

{\bf Subcase (ii).} $0\leq w(x,t)$ for all $x\in\mathbb{R}^N$ and $t\geq 0$, $\int_{\mathbb{R}^N}|w(x,t)^{\frac{1}{1-\theta}}dx<c'_w$ if $0<\theta<1$, we get
\begin{align}
&\quad\left(\int_0^{+\infty}\left(\int_{\mathbb{R}^N}w(x,t)[\Phi(V,u,W)]^{\theta}dx\right)^pdt\right)^{\frac{1}{p}}\nonumber\\
&\leq \left(\int_0^{+\infty}\left\{\left(\int_{\mathbb{R}^N}|w(x,t)^{\frac{1}{1-\theta}}dx\right)^{1-\vartheta}
\left(\int_{\mathbb{R}^N}\Phi(V,u,W)dx\right)^{\theta}\right\}^pdt\right)^{\frac{1}{p}} \nonumber\\
&\leq (c'_w)^{(1-\theta)}\left(\left(\int_0^1[2E(u_0)]dt+\int_1^{+\infty}\frac{C(u_0)}{4t^2}+\frac{4lE(u_0)+C(u_0)}{4t^{(2-l)}}\right)^{\theta p}dt\right)^{\frac{1}{p}}\nonumber\\
&\leq (c'_w)^{(1-\theta)}\tilde{c}'_1\left(\int_0^1[2E(u_0)]^{\theta p}dt\right)^{\frac{1}{p}}+(c'_w)^{(1-\theta)}\tilde{c}'_1\left(\int_1^{+\infty}\left(\frac{C(u_0)}{4t^2}+\frac{4lE(u_0)+C(u_0)}{4t^{(2-l)}}\right)^{\theta p}dt\right)^{\frac{1}{p}}\nonumber\\
&\leq [2E(u_0)]^{\theta}\tilde{c}'_1(c'_w)^{(1-\theta)}
+\frac{(c'_w)^{(1-\theta)}(\tilde{c}'_1)^2\tilde{c}'_2[C(u_0)]^{\theta}}{4^{\theta}}\left(\int_1^{+\infty}\frac{1}{t^{2\theta p}}dt\right)^{\frac{1}{p}}\nonumber\\
&\quad +
\frac{(c'_w)^{(1-\theta)}(\tilde{c}'_1)^2\tilde{c}'_2[4lE(u_0)+C(u_0)]^{\theta}}{4^{\theta}}\left(\int_1^{+\infty}\frac{1}{t^{(2-l)\theta p}}dt\right)^{\frac{1}{p}}\nonumber\\
&\leq (c'_w)^{(1-\theta)}\tilde{c}'_1\left([2E(u_0)]^{\theta}+\frac{\tilde{c}'_1\tilde{c}'_2[C(u_0)]^{\theta}}{4^{\theta}[2\theta p-1]^{\frac{1}{p}}}
+\frac{\tilde{c}'_1\tilde{c}'_2[4lE(u_0)+C(u_0)]^{\theta}}{4^{\theta}[(2-l)\theta p-1]^{\frac{1}{p}}}\right):=C'_1(u_0,p).\label{38x3}
\end{align}
Here $\tilde{c}'_1, \tilde{c}'_2=1$ if $\theta p<1$, $\tilde{c}'_1, \tilde{c}'_2=2^{\frac{\theta p-1}{\theta p}}$ if $\theta p\geq 1$.

{\bf Case (II).} Combined defocusing and focusing subcase in Case 2 of Theorem 3. In this case,
$$\int_{\mathbb{R}^N}\Phi(V,u,W)dx\leq \frac{2E(u_0)[1+C_r(u_0)]}{1-C_r(u_0)}$$
for $0\leq t\leq 1$ and
\begin{align}
\int_{\mathbb{R}^N}\Phi(V,u,W)dx\leq \frac{[1+C_r(u_0)]}{4[1-C_r(u_0)]}\left(\frac{C(u_0)}{t^2}+\frac{4lE(u_0)[1+C_r(u_0)]
+C(u_0)[1-C_r(u_0)]}{[1-C_r(u_0)]t^{2-\frac{l[1+C_r(u_0)]}{1-C_r(u_0)}}}\right)\label{38x4}
\end{align}
for $t\geq 1$.

Similarly, we get
\begin{align}
\left(\int_0^{+\infty}\left(\int_{\mathbb{R}^N}w(x,t)[\Phi(V,u,W)]^{\theta}dx\right)^pdt\right)^{\frac{1}{p}}\leq
C_2(u_0,p,\theta).
\end{align}
Here
\begin{align*}
C_2(u_0,p,\theta)=\frac{[2E(u_0)]^{\theta}[1+C_r(u_0)]^{\theta}\tilde{c}_w\tilde{c}_1}{[1-C_r(u_0)]^{\theta}}
+\frac{\tilde{c}_w\tilde{c}^2_1\tilde{c}'_2(C_1)^{\theta}}{(2\theta p-1)^{\frac{1}{p}}}+\tilde{c}_w\tilde{c}^2_1\tilde{c}'_2(C_2)^{\theta}(C_3)^{\frac{1}{p}}.
\end{align*}
and
\begin{align*}
&\tilde{c}_w=c_w\ {\rm if} \  \theta=1,\quad \tilde{c}_w=(c'_w)^{1-\theta} \ {\rm if} \  0<\theta<1,\\
&C_1=\frac{C(u_0)[1+C_r(u_0)]}{4[1-C_r(u_0)]},\\
&C_2=\frac{4lE(u_0)[1+C_r(u_0)]^2+C(u_0)[1-C^2_r(u_0)]}{4[1-C_r(u_0)]^2},\\
&C_3=\frac{[1-C_r(u_0)]}{(2\theta p-1)[1-C_r(u_0)]-l\theta p[1+C_r(u_0)]}.
\end{align*}

{\bf Case (III).} Case 1 of Theorem 3. In this case,
$$\int_{\mathbb{R}^N}\Phi(V,u,W)dx\leq C\quad {\rm for}\quad 0\leq t\leq 1,\qquad \int_{\mathbb{R}^N}\Phi(V,u,W)dx\leq \frac{C'}{t^2}\quad {\rm for} \quad t\geq 1. $$
Similarly, we have
\begin{align}
\left(\int_0^{+\infty}\left(\int_{\mathbb{R}^N}w(x,t)[\Phi(V,u,W)]^{\theta}dx\right)^pdt\right)^{\frac{1}{p}}\leq
C_3(u_0,p,\theta).
\end{align}
Here
\begin{equation}
\label{C(p)2}C_3(u_0,p,\theta)=\left\{
\begin{array}{lll}
\left([2E(u_0)]^{\theta}+\frac{[C(u_0)]^{\theta}}{4^{\theta}(2\theta p-1)^{\frac{1}{p}}}\right)(c'_w)^{(1-\theta)} \tilde{c}_1\quad {\rm in\ defocusing \ subcase}\\
\frac{(c'_w)^{(1-\theta)} \tilde{c}_1[1+C_r(u_0)]^{\theta}}{[1-C_r(u_0)]^{\theta}}\left([2E(u_0)]^{\theta}+\frac{[C(u_0)]^{\theta}}{4^{\theta}(2\theta p-1)^{\frac{1}{p}}}\right)\quad {\rm in\ combined }\\
\qquad \qquad \qquad  {\rm defocusing\ and\ focusing\ subcase}
\end{array}\right.
\end{equation}

{\bf Bound (G):}   Note that for $1\leq r\sigma<\gamma_2$, $1\leq r\sigma<\tilde{\gamma}_2$,
\begin{align}
\int_{\mathbb{R}^N}|G_1(|u|^2)|^{r\sigma}dx&\leq \left( m_3\|u_0\|^2_{L^2}\right)^{\frac{1}{\tau_3}}\left(m'_3 C_s\right)^{\frac{1}{\tau_4}}\left(\int_{\mathbb{R}^N}|\nabla h(|u|^2)|^2dx\right)^{\frac{2^*}{2\tau_4}}\nonumber\\
&\quad+\left( m_4\|u_0\|^2_{L^2}\right)^{\frac{1}{\tilde{\tau}_3}}\left(m'_4 C_s\right)^{\frac{1}{\tilde{\tau}_4}}\left(\int_{\mathbb{R}^N}|\nabla h(|u|^2)|^2dx\right)^{\frac{2^*}{2\tilde{\tau}_4}}\label{371}
\end{align}
and
\begin{align}
&\quad\left(\int_0^{+\infty}\left(\int_{\mathbb{R}^N}|G_1(|u|^2)|^{r\sigma}dx\right)^{\frac{q}{r\sigma}}dt\right)^{\frac{1}{q}}\nonumber\\
&\leq C_4(u_0,r,\sigma,\gamma_1,\gamma_2)\left(\int_0^{+\infty}\left(\int_{\mathbb{R}^N}\Phi(V,u,W)dx\right)^{\frac{2^*q}{2r\sigma\tau_4}}dt\right)^{\frac{1}{q}}\nonumber\\
&\quad+ \tilde{C}_4(u_0,r,\sigma,\tilde{\gamma}_1,\tilde{\gamma}_2)\left(\int_0^{+\infty}
\left(\int_{\mathbb{R}^N}\Phi(V,u,W)dx\right)^{\frac{2^*q}{2r\sigma\tilde{\tau}_4}}dt\right)^{\frac{1}{q}}. \label{36x3}
\end{align}
Here
\begin{align}
\frac{1}{\tau_3}=\frac{\gamma_2-r\sigma}{\gamma_2-\gamma_1},\quad \frac{1}{\tau_4}=\frac{r\sigma-\gamma_1}{\gamma_2-\gamma_1},\label{36x2}\\
\frac{1}{\tilde{\tau}_3}=\frac{\tilde{\gamma}_2-r\sigma}{\tilde{\gamma}_2-\tilde{\gamma}_1},\quad \frac{1}{\tilde{\tau}_4}=\frac{r\sigma-\tilde{\gamma}_1}{\tilde{\gamma}_2-\tilde{\gamma}_1}.\label{327x1'}
\end{align}
and $\tilde{c}_3=1$ if $q\leq r\sigma$, $\tilde{c}_3=2^{\frac{q-r\sigma}{r\sigma}}$ if $q>r\sigma$, $\tilde{c}_4=1$ if $q>1$, $\tilde{c}_4=2^{\frac{1-q}{q}}$ if $q\leq 1$,
\begin{align}
C(u_0,r,\sigma,\gamma_1,\gamma_2)&=\left(m_3\|u_0|^2_{L^2}\right)^{\frac{1}{r\sigma\tau_3}}\left(m'_3 C_s\right)^{\frac{1}{r\sigma\tau_4}},\label{5271}\\
C_4(u_0,r,\sigma,\gamma_1,\gamma_2)&=C(u_0,r,\sigma,\gamma_1,\gamma_2)\tilde{c}^{\frac{1}{q}}_3\tilde{c}_4\label{327w1}\\
\tilde{C}(u_0,r,\sigma, \tilde{\gamma}_1,\tilde{\gamma}_2)&=\left( m_4\|u_0|^2_{L^2}\right)^{\frac{1}{r\sigma\tilde{\tau}_3}}\left(m'_4 C_s\right)^{\frac{1}{r\sigma\tilde{\tau}_4}},\label{5272}\\ \tilde{C}_4(u_0,r,\sigma,\tilde{\gamma}_1,\tilde{\gamma}_2)&=\tilde{C}(u_0,r,\sigma,\tilde{\gamma}_1,\tilde{\gamma}_2)\tilde{c}^{\frac{1}{q}}_3\tilde{c}_4.\label{36x4}
\end{align}

We prove it in two cases.

 Case (IV). Case 1 in combined defocusing and focusing case of Theorem 3;

Case (V). Case 2 in combined defocusing and focusing case of Theorem 3.

We only give the details in Case (V), the proof in Case (IV) is similar.

{\bf Case (V).}  Denote $$\tilde{C}_1=1\quad {\rm if}\quad 2^*q\leq 2r\sigma\tau_4,\quad \tilde{C}_1=2^{\frac{2^*q-2r\sigma\tau_4}{2r\tau_4}},\quad {\rm if} 2^*q>2r\sigma\tau_4;$$ $$\tilde{C}_2=1\quad {\rm if}\quad 2^*q\leq 2r\tilde{\tau}_4, \quad \tilde{C}_2=2^{\frac{2^*q-2r\sigma\tilde{\tau}_4}{2r\sigma\tilde{\tau}_4}}\quad {\rm if}\quad 2^*q>2r\sigma\tilde{\tau}_4.$$
 And
\begin{align}
C_5(u_0,r,\sigma,\gamma_1,\gamma_2)=C_4(u_0,r,\sigma,\gamma_1,\gamma_2)\tilde{c}_4\left(\frac{1+C_r(u_0)}{1-C_r(u_0)}\right)^{\frac{2^*}{2r\sigma\tau_4}},\\
C_6(u_0,r,\sigma,\tilde{\gamma}_1,\tilde{\gamma}_2)=\tilde{C}_4(u_0,r,\sigma,\tilde{\gamma}_1,\tilde{\gamma}_2)\tilde{c}_4
\left(\frac{1+C_r(u_0)}{1-C_r(u_0)}\right)^{\frac{2^*}{2r\sigma\tilde{\tau}_4}}.
\end{align}

We also discuss it in two subcases.

{\bf Subcase (i).} $0\leq w(x,t)\leq c_w$ for any $x\in\mathbb{R}^N$ and $t\geq 0$, $1\leq r<\min(\gamma_2, \tilde{\gamma}_2)$. Taking $\sigma=1$ in (\ref{36x3}),  we obtain
\begin{align}
&\quad\left(\int_0^{+\infty}\left(\int_{\mathbb{R}^N}w(x,t)|G_1(|u|^2)|^rdx\right)^{\frac{q}{r}}dt\right)^{\frac{1}{q}}\nonumber\\
&\leq (c_w)^{\frac{1}{r}}C_5(u_0,r,\gamma_1,\gamma_2)\left\{\left(\int_0^1[2E(u_0)]^{\frac{2^*q}{2r\tau_4}}dt\right)^{\frac{1}{q}}+\right.\nonumber\\
&\qquad\left.\left(\int_1^{+\infty} \left(\frac{C(u_0)}{4t^2}+\frac{4lE(u_0)[1+C_r(u_0)]+C(u_0)[1-C_r(u_0)]}{4[1-C_r(u_0)]t^{2-\frac{l[1+C_r(u_0)]}{1-C_r(u_0)}}}
\right)^{\frac{2^*q}{2r\tau_4}}dt \right)^{\frac{1}{q}}\right\}\nonumber\\
&\quad+(c_w)^{\frac{1}{r}}C_6(u_0,r,\tilde{\gamma}_1,\tilde{\gamma}_2)
\left\{\left(\int_0^1[2E(u_0)]^{\frac{2^*q}{2r\tilde{\tau}_4}}dt\right)^{\frac{1}{q}}+\right.\nonumber\\
&\qquad\left.\left(\int_1^{+\infty} \left(\frac{C(u_0)}{4t^2}+\frac{4lE(u_0)[1+C_r(u_0)]+C(u_0)[1-C_r(u_0)]}{4[1-C_r(u_0)]
t^{2-\frac{l[1+C_r(u_0)]}{1-C_r(u_0)}}}\right)^{\frac{2^*q}{2r\tilde{\tau}_4}}dt \right)^{\frac{1}{q}}\right\}\nonumber\\
&\leq (c_w)^{\frac{1}{r}}C_5(u_0,r,\gamma_1,\gamma_2)\left\{[2E(u_0)]^{\frac{2^*}{2r\tau_4}}
+\tilde{C}_1^{\frac{1}{q}}\tilde{c}_4\left(\frac{C(u_0)}{4}\right)^{\frac{2^*}{2r\tau_4}}\left(\frac{r\tau_4}{2^*q-r\tau_4}   \right)^{\frac{1}{q}}\right.\nonumber\\
&\left.+\tilde{C}_1^{\frac{1}{q}}\tilde{c}_4\left(\frac{4lE(u_0)[1+C_r(u_0)]+C(u_0)[1-C_r(u_0)]}{4[1-C_r(u_0)]}\right)^{\frac{2^*}{2r\tau_4}}
\left(\int_1^{+\infty} \frac{1}{t^{\frac{2^*q}{2r\tau_4}[2-\frac{l[1+C_r(u_0)]}{1-C_r(u_0)}]}}dt \right)^{\frac{1}{q}}\right\}\nonumber\\
&\quad+(c_w)^{\frac{1}{r}}C_6(u_0,r,\tilde{\gamma}_1,\tilde{\gamma}_2)\left\{[2E(u_0)]^{\frac{2^*}{2r\tilde{\tau}_4}}
+\tilde{C}_2^{\frac{1}{q}}\tilde{c}_4\left(\frac{C(u_0)}{4}\right)^{\frac{2^*}{2r\tilde{\tau}_4}}\left(\frac{r\tilde{\tau}_4}{2^*q-r\tilde{\tau}_4}   \right)^{\frac{1}{q}}\right.\nonumber\\
&\left.+\tilde{C}_2^{\frac{1}{q}}\tilde{c}_4\left(\frac{4lE(u_0)[1+C_r(u_0)]+C(u_0)[1-C_r(u_0)]}{4[1-C_r(u_0)]}\right)^{\frac{2^*}{2r\tilde{\tau}_4}}
\left(\int_1^{+\infty} \frac{1}{t^{\frac{2^*q}{2r\tilde{\tau}_4}[2-\frac{l[1+C_r(u_0)]}{1-C_r(u_0)}]}}dt \right)^{\frac{1}{q}}\right\}\nonumber\displaybreak\\
&=(c_w)^{\frac{1}{r}}C_5(u_0,r,\gamma_1,\gamma_2)\left([2E(u_0)]^{\frac{2^*}{2r\tau_4}}
+\tilde{C}_1^{\frac{1}{q}}\tilde{c}_4\left(\frac{C(u_0)}{4}\right)^{\frac{2^*}{2r\tau_4}}
\left(\frac{r\tau_4}{2^*q-r\tau_4}   \right)^{\frac{1}{q}}\right)\nonumber\\
&\quad+(c_w)^{\frac{1}{r}}C_5(u_0,r,\gamma_1,\gamma_2)\tilde{C}_1^{\frac{1}{q}}\tilde{c}_4
\left(\frac{4lE(u_0)[1+C_r(u_0)]+C(u_0)[1-C_r(u_0)]}{4[1-C_r(u_0)]}\right)^{\frac{2^*}{2r\tau_4}}\nonumber\\
&\qquad \times \left(\frac{2r\tau_4[1-C_r(u_0)]}{(22^*q-2r\tau_4)[1-C_r(u_0)]-2^*ql[1+C_r(u_0)]} \right)^{\frac{1}{q}}\nonumber\\
&+\quad (c_w)^{\frac{1}{r}}C_6(u_0,r,\tilde{\gamma}_1,\tilde{\gamma}_2)\left([2E(u_0)]^{\frac{2^*}{2r\tilde{\tau}_4}}
+\tilde{C}_2^{\frac{1}{q}}\tilde{c}_4\left(\frac{C(u_0)}{4}\right)^{\frac{2^*}{2r\tilde{\tau}_4}}\left(\frac{r\tilde{\tau}_4}{2^*q-r\tilde{\tau}_4}   \right)^{\frac{1}{q}}\right)\nonumber\\
&\quad+(c_w)^{\frac{1}{r}}C_6(u_0,r,\tilde{\gamma}_1,\tilde{\gamma}_2)\tilde{C}_2^{\frac{1}{q}}\tilde{c}_4
\left(\frac{4lE(u_0)[1+C_r(u_0)]+C(u_0)[1-C_r(u_0)]}{4[1-C_r(u_0)]}\right)^{\frac{2^*}{2r\tilde{\tau}_4}}\nonumber\\
&\qquad \times \left(\frac{2r\tilde{\tau}_4[1-C_r(u_0)]}{(22^*q-2r\tilde{\tau}_4)[1-C_r(u_0)]-2^*ql[1+C_r(u_0)]}\right)^{\frac{1}{q}}.\label{37C3}
\end{align}

{\bf Subcase (ii).} $0\leq w(x,t)$ for any $x\in\mathbb{R}^N$ and $t\geq 0$, $\int_{\mathbb{R}^N}|w(x,t)|^{\frac{\delta}{\delta-1}}dx\leq c'_w$ for some $1<\delta<\frac{\gamma_2}{r}\leq\gamma_2$ and $1<\delta<\frac{\tilde{\gamma}_2}{r}\leq\tilde{\gamma}_2$. Taking $\sigma=\delta$ in (\ref{36x3}),
 we have
\begin{align}
&\quad\left(\int_0^{+\infty}\left(\int_{\mathbb{R}^N}v(x,t)|G_1(|u|^2)|^rdx\right)^{\frac{q}{r}}dt\right)^{\frac{1}{q}}\nonumber\\
&\leq \left(\int_0^{+\infty}\left\{\left(\int_{\mathbb{R}^N}|v(x,t)|^{\frac{\delta}{\delta-1}}dx\right)^{\frac{\delta-1}{\delta}}
\left(\int_{\mathbb{R}^N}|G_1(|u|^2)|^{r\delta}dx\right)^{\frac{1}{\delta}}\right\}^{\frac{q}{r}}dt\right)^{\frac{1}{q}}\nonumber\\
&\leq (c'_v)^{\frac{\delta-1}{r\delta}}C_5\left\{\left(\int_0^1[2E(u_0)]^{\frac{2^*q}{2r\delta\tau_4}}dt\right)^{\frac{1}{q}}+\right.\nonumber\\
&\qquad\left.\left(\int_1^{+\infty} \left(\frac{C(u_0)}{4t^2}+\frac{4lE(u_0)[1+C_r(u_0)]+C(u_0)[1-C_r(u_0)]}{4[1-C_r(u_0)]t^{2-\frac{l[1+C_r(u_0)]}{1-C_r(u_0)}}}
\right)^{\frac{2^*q}{2r\delta\tau_4}}dt \right)^{\frac{1}{q}}\right\}\nonumber\\
&\quad+(c'_v)^{\frac{\delta-1}{r\delta}}C_6\left\{\left(\int_0^1[2E(u_0)]^{\frac{2^*q}{2r\delta\tilde{\tau}_4}}dt\right)^{\frac{1}{q}}+\right.\nonumber\\
&\qquad\left.\left(\int_1^{+\infty} \left(\frac{C(u_0)}{4t^2}
+\frac{4lE(u_0)[1+C_r(u_0)]+C(u_0)[1-C_r(u_0)]}{4[1-C_r(u_0)]t^{2-\frac{l[1+C_r(u_0)]}{1-C_r(u_0)}}}\right)^{\frac{2^*q}{2r\delta\tilde{\tau}_4}}dt \right)^{\frac{1}{q}}\right\}\nonumber\displaybreak\\
&\leq (c'_v)^{\frac{\delta-1}{r\delta}}C_5\left([2E(u_0)]^{\frac{2^*}{2r\delta\tau_4}}
+\tilde{C}_1^{\frac{1}{q}}\tilde{c}_4\left(\frac{C(u_0)}{4}\right)^{\frac{2^*}{2r\delta\tau_4}}\left(\frac{r\delta\tau_4}{2^*q-r\delta\tau_4}   \right)^{\frac{1}{q}}\right)\nonumber\\
&\quad+(c'_v)^{\frac{\delta-1}{r\delta}}C_5\tilde{C}_1^{\frac{1}{q}}\tilde{c}_4
\left(\frac{4lE(u_0)[1+C_r(u_0)]+C(u_0)[1-C_r(u_0)]}{4[1-C_r(u_0)]}\right)^{\frac{2^*}{2r\delta\tau_4}}\nonumber\\
&\qquad \times \left(\frac{2r\delta\tau_4[1-C_r(u_0)]}{(22^*q-2r\delta\tau_4)[1-C_r(u_0)]-2^*ql[1+C_r(u_0)]} \right)^{\frac{1}{q}}\nonumber\\
&+\quad (c'_v)^{\frac{\delta-1}{r\delta}}C_6\left([2E(u_0)]^{\frac{2^*}{2r\delta\tilde{\tau}_4}}
+\tilde{C}_2^{\frac{1}{q}}\tilde{c}_4\left(\frac{C(u_0)}{4}\right)^{\frac{2^*}{2r\delta\tilde{\tau}_4}}
\left(\frac{r\delta\tilde{\tau}_4}{2^*q-r\delta\tilde{\tau}_4}   \right)^{\frac{1}{q}}\right)\nonumber\\
&\quad+(c'_v)^{\frac{\delta-1}{r\delta}}C_6\tilde{C}_2^{\frac{1}{q}}\tilde{c}_4
\left(\frac{4lE(u_0)[1+C_r(u_0)]+C(u_0)[1-C_r(u_0)]}{4[1-C_r(u_0)]}\right)^{\frac{2^*}{2r\delta\tilde{\tau}_4}}\nonumber\\
&\qquad \times \left(\frac{2r\delta\tilde{\tau}_4[1-C_r(u_0)]}{(22^*q-2r\delta\tilde{\tau}_4)[1-C_r(u_0)]-2^*ql[1+C_r(u_0)]}\right)^{\frac{1}{q}}.\label{37C3}
\end{align}

As a corollary of Theorem 3 and Theorem 4, we can obtain the  decay rate and asymptotic behavior for the solution as $t\rightarrow +\infty$.

{\bf Corollary 4.1.} {\it Let $u(x,t)$ be the global solution of (\ref{1}). Under the assumptions of Theorem 3 and Theorem 4,
\begin{align}
&\int_{\mathbb{R}^N}[\delta_h|\nabla h(|u|^2)|^2+|V(x)||u|^2+|G_1(|u|^2)|+|G_2(|u|^2)|+\frac{1}{2}(W*|u|^2)|u|^2]dx\leq \frac{C}{t^2}\label{681}
\end{align}
in Case 1,
\begin{align}
&\int_{\mathbb{R}^N}[\delta_h|\nabla h(|u|^2)|^2+|V(x)||u|^2+|G_2(|u|^2)|+\frac{1}{2}(W*|u|^2)|u|^2]dx\leq \frac{C}{t^{2-l}}\label{682'}
\end{align}
in defocusing subcase of Case 2,
\begin{align}
&\int_{\mathbb{R}^N}[\delta_h|\nabla h(|u|^2)|^2+|V(x)||u|^2+|G_1(|u|^2)|+|G_2(|u|^2)|+\frac{1}{2}(W*|u|^2)|u|^2]dx\nonumber\\
&\leq \frac{C}{t^{2-\frac{l(1+C_r(u_0))}{1-C_r(u_0)}}}\label{682}
\end{align}
in combined defocusing and focusing subcase of Case 2, and
\begin{align}
&\lim_{t\rightarrow +\infty}\int_{\mathbb{R}^N}[\delta_h|\nabla h(|u|^2)|^2+|V(x)||u|^2+|G(|u|^2)|+\frac{1}{2}(W*|u|^2)|u|^2]dx=0,\label{641}\\
&\lim_{t\rightarrow +\infty} \int_{\mathbb{R}^N}|\nabla u|^2dx=2E(u_0),\quad \lim_{t\rightarrow +\infty} \int_{\mathbb{R}^N}[|u|^2+|\nabla u|^2]dx=M(u_0)+2E(u_0).\label{642}
\end{align}

Consequently, for any $2\leq r<2^*$, $2^*=\frac{2N}{N-2}$ if $N\geq 3$, $2^*=+\infty$ if $N=1,2$,
\begin{align}
\int_{\mathbb{R}^N}|u|^rdx\leq C.\label{8201}
\end{align}
}

{\bf Proof of Corollary 4.1:} (\ref{681}),(\ref{682'}, (\ref{682}) and (\ref{641}) are the direct results of (\ref{222x2}), (\ref{222x2'}), (\ref{222x5}) and (\ref{222x2''}).

By mass and energy conservation laws, we have
\begin{align*}
\frac{1}{2}\int_{\mathbb{R}^N}|\nabla u|^2dx=E(u_0)-\frac{1}{2}\int_{\mathbb{R}^N}[\delta_h|\nabla h(|u|^2)|^2+|V(x)||u|^2+|G(|u|^2)|+\frac{1}{2}(W*|u|^2)|u|^2]dx,
\end{align*}
which means (\ref{642}). (\ref{641}) and (\ref{642}) imply that
$$\int_{\mathbb{R}^N}|u|^2dx\leq C, \quad \int_{\mathbb{R}^N}|\nabla u|^2dx\leq C,$$
by embedding theorem, we get (\ref{8201}).\hfill $\Box$

We give two examples to show the results on Theorem 3 and Theorem 4.

{\bf Remark 4.1.} 1. If $h(s)\equiv0$, $F(|u|^2)=-|u|^{2\beta}$, $V(x)=-\frac{1}{|x|^m}$ and $W(x)=-\frac{1}{|x|^n}$, $x\neq \mathbf{0}$, $\alpha, \beta, m, n>0$, then we especially have
\begin{align}
&\int_0^{\infty}\int_{\mathbb{R}^N}\left[\frac{|u|^{2\beta+2}}{\beta+1}+\frac{1}{|x|^m}|u|^2+\frac{1}{2}(\frac{1}{|x|^n}*|u|^2)|u|^2\right]dxdt\leq C\label{224w2}\\
&\int_0^{\infty}\left(\int_{\mathbb{R}^N}\left[\frac{|u|^{2\beta+2}}{\beta+1}+\frac{1}{|x|^m}|u|^2+\frac{1}{2}(\frac{1}{|x|^n}*|u|^2)|u|^2\right]dx\right)^pdt\leq C\\
&\|u\|_{L^q(\mathbb{R}^+)L^{\tilde{r}}(\mathbb{R}^N)}=\left(\int_0^{\infty}\left(\int_{\mathbb{R}^N}|u|^{\tilde{r}}dx\right)^{\frac{q}{\tilde{r}}}dt\right)^{\frac{1}{q}}
\leq C
\end{align}
under certain assumptions.

2. Consider the following Cauchy problem:
\begin{equation}
\label{632} \left\{
\begin{array}{lll}
iu_t=\Delta u+2\alpha |u|^{2\alpha-2}u\Delta (|u|^{2\alpha})-\frac{|x|^2u}{|x|^2+1}\mp|u|^{2\beta}u-(\frac{|x|^2}{(a|x|^2+1)^m}*|u|^2)u, x\in \mathbb{R}^N, t>0\\
u(x,0)=u_0(x),\quad x\in \mathbb{R}^N.
\end{array}\right.
\end{equation}
Here $\alpha\geq \frac{1}{2}$.
Then
\begin{align*}
&2h''(s)h'(s)s+(h'(s))^2=(2\alpha-1)\alpha^2s^{2\alpha-2},\\
& NF(s)s-(N+2)G(s)=\mp[N-\frac{N+2}{\beta+1}]s^{\beta+1},\\
&2V+(x\cdot \nabla V)\leq 0,\quad 2W+(x\cdot \nabla W)\leq 0 \quad {\rm for  \ suitable}\quad a,\  m.
\end{align*}
Then
$$
k_1=2\alpha-1,\quad k_2=\frac{|N\beta-2|}{\beta+1},\quad l=\max(k_1, k_2),
$$
\begin{align*}
&\int_0^{\infty}\int_{\mathbb{R}^N}\left[|\nabla h(|u|^2)|^2+\frac{|u|^{2\beta+2}}{\beta+1}+|V(x)||u|^2+\frac{1}{2}(|W(x)|*|u|^2)|u|^2\right]dxdt\leq C,\\
&\int_0^{\infty}\left(\int_{\mathbb{R}^N}\left[|\nabla h(|u|^2)|^2+\frac{|u|^{2\beta+2}}{\beta+1}+|V(x)||u|^2+\frac{1}{2}(|W(x)|*|u|^2)|u|^2\right]dx\right)^pdt\leq C,\\
&\|u\|_{L^q(\mathbb{R}^+)L^{\tilde{r}}(\mathbb{R}^N)}=\left(\int_0^{\infty}\left(\int_{\mathbb{R}^N}|u|^{\tilde{r}}dx\right)^{\frac{q}{\tilde{r}}}dt\right)^{\frac{1}{q}}
\leq C
\end{align*}
under certain assumptions.

We require the assumption of $xu_0\in L^2(\mathbb{R}^N)$ in Theorem 3 and Theorem 4. However, we can remove the restriction $xu_0\in L^2(\mathbb{R}^N)$ if we use interaction Morawetz estimates for the solution belonging $H^{\frac{1}{2}}(\mathbb{R}^N)\cap X$ in the next section.

\section{Interaction Morawetz inequality}
\qquad  We assume that
\begin{align}
&G(s)-F(s)s\geq 0,\quad [h'(s)]^2+2h''(s)h'(s)s\geq 0\quad  {\rm for}\ s\geq 0,\label{352}\\
&x\cdot\nabla V(x)\geq 0, \quad x\cdot \nabla W(x)\geq 0\quad {\rm for}\ x\in \mathbb{R}^N\quad {\rm if}\quad N\geq 2.\label{353}
\end{align}

\subsection{Interaction Morawetz inequality in dimension $N\geq 3$}
\qquad Let
\begin{align}
M_a^{\otimes_2}(t)=2\int_{\mathbb{R}^N\otimes\mathbb{R}^N}\nabla a(x-y)\cdot \Im \left[(u(x,t)u(y,t))\nabla (\bar{u}(x,t)\bar{u}(y,t))\right]dxdy,\label{2171}
\end{align}
where $\nabla=(\partial_x, \partial_y)$ and
\begin{align}
\tilde{H}(|u|^2)=\int_0^{|u|^2}[h'(s)]^2sds.\label{351}
\end{align}
Then
\begin{align}
\frac{d}{dt}M_a^{\otimes_2}(t)&=-\int_{\mathbb{R}^N\otimes\mathbb{R}^N}[4\tilde{H}(|u(x,t)|^2)+|u(x,t)|^2]\Delta_x(\Delta_x a(x,y))dx|u(y,t)|^2dy\nonumber\\
&\quad +4\int_{\mathbb{R}^N\otimes\mathbb{R}^N}\sum_{j,k=1}^N\Re(\partial_{x_k}u(x,t)\partial_{x_j}\bar{u}(x,t))\partial_{x_k}\partial_{x_j}a(x,y)dx|u(y,t)|^2dy\nonumber
\\
&\quad +4\int_{\mathbb{R}^N\otimes\mathbb{R}^N}\sum_{j,k=1}^N\partial_{x_k}h(|u(x,t)|^2)\partial_{x_j}h(|u(x,t)|^2)\partial_{x_k}\partial_{x_j}a(x,y)dx
|u(y,t)|^2dy\nonumber\\
&\quad +2\int_{\mathbb{R}^N\otimes\mathbb{R}^N}h'(|u|^2)[h'(|u|^2)+2h''(|u|^2)|u|^2]|\nabla_x(|u(x,t)|^2)|^2\Delta_xa(x,y)dx|u(y,t)|^2dy\nonumber\displaybreak\\
&\quad +2\int_{\mathbb{R}^N\otimes\mathbb{R}^N}[G(|u(x,t)|^2)-F(|u(x,t)|^2)|u(x,t)|^2]\Delta_xa(x,y)dx|u(y,t)|^2dy\nonumber\\
&\quad +2\int_{\mathbb{R}^N\otimes\mathbb{R}^N}(\nabla_x V(x)\cdot \nabla_x a(x,y))|u(x,t)|^2dx|u(y,t)|^2dy\nonumber\\
&\quad +\int_{\mathbb{R}^N\otimes\mathbb{R}^N}[(\nabla_x W(x)\cdot \nabla_x a(x,y))*|u(x,t)|^2]|u(x,t)|^2dx|u(y,t)|^2dy\nonumber\\
&\quad-\int_{\mathbb{R}^N\otimes\mathbb{R}^N}[4\tilde{H}(|u(y,t)|^2)+|u(y,t)|^2]\Delta_y(\Delta_y a(x,y))dy|u(x,t)|^2dx\nonumber\\
&\quad +4\int_{\mathbb{R}^N\otimes\mathbb{R}^N}\sum_{j,k=1}^N\Re(\partial_{y_k}u(y,t)\partial_{y_j}\bar{u}(y,t))
\partial_{y_k}\partial_{y_j}a(x,y)dy|u(x,t)|^2dx\nonumber\\
&\quad +4\int_{\mathbb{R}^N\otimes\mathbb{R}^N}\sum_{j,k=1}^N\partial_{y_k}h(|u(y,t)|^2)\partial_{y_j}h(|u(y,t)|^2)
\partial_{y_k}\partial_{y_j}a(x,y)dy|u(x,t)|^2dx\nonumber\\
&\quad +2\int_{\mathbb{R}^N\otimes\mathbb{R}^N}h'(|u|^2)[h'(|u|^2)+2h''(|u|^2)|u|^2]|\nabla_y(|u(y,t)|^2)|^2\Delta_ya(x,y)dy|u(x,t)|^2dx\nonumber\\
&\quad +2\int_{\mathbb{R}^N\otimes\mathbb{R}^N}[G(|u(y,t)|^2)-F(|u(y,t)|^2)|u(y,t)|^2]\Delta_ya(x-y)dy|u(x,t)|^2dx\nonumber\\
&\quad +2\int_{\mathbb{R}^N\otimes\mathbb{R}^N}(\nabla_y V(y)\cdot \nabla_y a(x,y))|u(y,t)|^2dy|u(x,t)|^2dx\nonumber\\
&\quad +\int_{\mathbb{R}^N\otimes\mathbb{R}^N}[(\nabla_y W(y)\cdot \nabla_y a(x,y))*|u(y,t)|^2]|u(y,t)|^2dy|u(x,t)|^2dx.\label{2172}
\end{align}

If $a(x,y)=|x-y|$, then $a(x,y)$ is convex with respect to both $x$ and $y$, and
\begin{equation}
\label{2173}-\Delta_x\Delta_x a(x,y)=-\Delta_y\Delta_y a(x,y)\left\{
\begin{array}{lll}
8\pi\delta(x-y)\quad {\rm if} \  N=3,\\
\frac{(N-1)(N-3)}{|x-y|^3}\quad {\rm if} \ N\geq 4.
\end{array}\right.
\end{equation}
Here $\delta(x-y)$ is Dirac function
\begin{equation*}
\delta(x-y)=\left\{
\begin{array}{lll}
\infty\quad& {\rm if} \  x-y=\mathbf{0},\\
0\quad &{\rm if} \ x-y\neq \mathbf{0}.
\end{array}\right.
\end{equation*}

Under the assumptions on $h(s)$, $F(s)$, $V(x)$ and $W(x)$, we get
\begin{align}
\frac{d}{dt}M_a^{\otimes_2}(t)&\geq-\int_{\mathbb{R}^N\otimes\mathbb{R}^N}[4\tilde{H}(|u(x,t)|^2)+|u(x,t)|^2]\Delta_x(\Delta_x a(x,y))dx|u(y,t)|^2dy\nonumber\\
&\quad-\int_{\mathbb{R}^N\otimes\mathbb{R}^N}[4\tilde{H}(|u(y,t)|^2)+|u(y,t)|^2]\Delta_y(\Delta_y a(x,y))dy|u(x,t)|^2dx.\label{218x1}
\end{align}

Therefore, if $N=3$, by the property of the function $\delta(x-y)$, we have
\begin{align}
\int_0^T\int_{\mathbb{R}^3}[|u(x,t)|^4+\tilde{H}(|u(x,t)|^2)|u(x,t)|^2]dxdt\leq C\sup_{t\in[0,T]}|M_a^{\otimes_2}(t)|.\label{2174}
\end{align}
If $N\geq 4$, we obtain
\begin{align}
&\quad \int_0^T\int_{\mathbb{R}^N\otimes\mathbb{R}^N}[\frac{|u(x,t)|^2|u(y,t)|^2}{|x-y|^3}+
\frac{\sqrt{|u(x,t)|^2\tilde{H}(|u(x,t)|^2)}\sqrt{|u(y,t)|^2\tilde{H}(|u(y,t)|^2)}}{|x-y|^3}]dxdydt\nonumber\\
&\leq \int_0^T\int_{\mathbb{R}^N\otimes\mathbb{R}^N}[\frac{|u(x,t)|^2|u(y,t)|^2}{|x-y|^3}+
\frac{|u(y,t)|^2\tilde{H}(|u(x,t)|^2)+|u(x,t)|^2\tilde{H}(|u(y,t)|^2)}{|x-y|^3}]dxdydt\nonumber\\
&\leq C\sup_{t\in[0,T]}|M_a^{\otimes_2}(t)|.\label{2175}
\end{align}
Similar to the proof of Theorem 2.17 in \cite{Colliander5}, using Plancherel theorem, we get
\begin{align}
\int_{\mathbb{R}^N\otimes\mathbb{R}^N}\frac{|u(x,t)|^2|u(y,t)|^2}{|x-y|^3}dxdy\simeq \int_{\mathbb{R}^N}|D^{-\frac{N-3}{2}}(|u(x)|^2)|^2dx\label{2176}
\end{align}
and
\begin{align}
&\quad\int_{\mathbb{R}^N\otimes\mathbb{R}^N}\frac{\sqrt{|u(x,t)|^2\tilde{H}(|u(x,t)|^2)}\sqrt{|u(y,t)|^2\tilde{H}(|u(y,t)|^2)}}{|x-y|^3}dxdy\nonumber\\
&\simeq \int_{\mathbb{R}^N}\left|D^{-\frac{N-3}{2}}\left(\sqrt{|u(x,t)|^2\tilde{H}(|u(x,t)|^2)}\right)\right|^2dx.\label{2177}
\end{align}
By the results of \cite{Colliander1, Colliander4, Colliander5} and using mass conservation law, we have
\begin{align}
\sup_{t\in[0,T]}|M_a^{\otimes_2}(t)|\leq C'\sup_{t\in[0,T]}\|u(x,t)\|_{\dot{H}^{1/2}_x}^2\|u\|_{L^2_x}^2\leq C\sup_{t\in[0,T]}\|u(x,t)\|_{\dot{H}^{1/2}_x}^2.\label{2178}
\end{align}
Combining (\ref{2174})--(\ref{2178}), we have
\begin{align}
\|D^{-\frac{N-3}{2}}\left(|u(x,t)|^2+\sqrt{|u(x,t)|^2\tilde{H}(|u(x,t)|^2)}\right)\|_{L^2_tL^2_x}^2\leq C\sup_{t\in[0,T]}\|u(x,t)\|_{\dot{H}^{1/2}_x}^2,\label{2179}
\end{align}
which is the interaction Morawetz estimates for (\ref{1}) in the case of $N\geq 3$.

\subsection{Interaction Morawetz inequality in dimension $N=2$}
\qquad Inspired by \cite{Colliander1, Colliander4, Colliander5, Selvitella}, we will choose $a(x,y)=a(|x-y|)$ in the Morawetz action when $N=2$, where $a(r)$ is a radial function satisfies
\begin{align}
\Delta a(r)=\int^{\infty}_r s\log(\frac{s}{r})w_{r_0}(s)ds,\label{21710}
\end{align}
where $r_0$ is a small positive number and
\begin{equation}
\label{218x2}w_{r_0}(s):=\left\{
\begin{array}{lll}
\frac{1}{s^3}\quad & {\rm if} \  s\geq r_0,\cr\\
0 \quad\quad & {\rm otherwise}.
\end{array}\right.
\end{equation}

Let
\begin{align}
M_a^{\otimes_2}(t)=2\int_{\mathbb{R}^2\otimes\mathbb{R}^2}\nabla a(|x-y|)\cdot \Im \left(\overline{u(x,t)u(y,t)}\nabla (u(x,t)u(y,t))\right)dxdy.\label{218x3}
\end{align}
Similar to (\ref{2172}) and (\ref{218x1}), and using
$$
-\Delta \Delta a(|x|)=\frac{2\pi}{r_0}\delta(|x|)-w_{r_0}(|x|),
$$
we can get
\begin{align}
&\quad \int_0^T\int_{\mathbb{R}^2}\frac{2\pi}{r_0}[|u(x,t)|^4+4\tilde{H}(|u(x,t)|^2)|u(x,t)|^2]dxdt\nonumber\\
&+ \int_0^T\int_{\mathbb{R}^2}\frac{2\pi}{r_0}[|u(y,t)|^4+4\tilde{H}(|u(y,t)|^2)|u(y,t)|^2]dydt\nonumber\\
&-2\int_0^T\int_{\mathbb{R}^2\otimes\mathbb{R}^2}w_{r_0}(|x-y|)|u(x,t)|^2|u(y,t)|^2dxdydt\nonumber\\
&-4\int_0^T\int_{\mathbb{R}^2\otimes\mathbb{R}^2}w_{r_0}(|x-y|)\tilde{H}(|u(x,t)|^2)|u(y,t)|^2dxdydt\nonumber\\
&-4\int_0^T\int_{\mathbb{R}^2\otimes\mathbb{R}^2}w_{r_0}(|x-y|)\tilde{H}(|u(y,t)|^2)|u(x,t)|^2dxdydt\nonumber\\
&\leq C\sup_{t\in[0,T]}|M_a^{\otimes_2}(t)|.\label{218x5}
\end{align}
Noticing that
\begin{align}
\int_{\mathbb{R}^2}w_{r_0}(|x-y|)dx=\int_{\mathbb{R}^2}w_{r_0}(|x-y|)dy=\frac{2\pi}{r_0},\label{218x6}
\end{align}
using (\ref{218x5}), we have
\begin{align}
&\quad\int_0^T\int_{\mathbb{R}^2\otimes\mathbb{R}^2}w_{r_0}(|x-y|)[|u(x,t)|^2-|u(y,t)|^2]^2dxdydt\nonumber\\
&+4\int_0^T\int_{\mathbb{R}^2\otimes\mathbb{R}^2}w_{r_0}(|x-y|)[|u(x,t)|^2-|u(y,t)|^2][\tilde{H}(|u(x,t)|^2)-\tilde{H}(|u(y,t)|^2)]dxdydt\nonumber\\
&\leq C'\sup_{t\in[0,T]}|M_a^{\otimes_2}(t)|.\label{218x7}
\end{align}
Let $r_0\rightarrow 0$, we get
\begin{align}
&\quad\int_0^T\int_{\mathbb{R}^2\otimes\mathbb{R}^2}\frac{[|u(x,t)|^2-|u(y,t)|^2]^2}{|x-y|^3}dxdydt\nonumber\\
&+\int_0^T\int_{\mathbb{R}^2\otimes\mathbb{R}^2}\frac{[|u(x,t)|^2-|u(y,t)|^2][\tilde{H}(|u(x,t)|^2)-\tilde{H}(|u(y,t)|^2)]}{|x-y|^3}dxdydt\nonumber\\
&\leq C'\sup_{t\in[0,T]}|M_a^{\otimes_2}(t)|\leq C\sup_{t\in[0,T]}\|u(x,t)\|^2_{\dot{H}^{\frac{1}{2}}_x}.\label{218x7}
\end{align}
Note that
$$
\int_{\mathbb{R}^2\otimes\mathbb{R}^2}\frac{[|u(x,t)|^2-|u(y,t)|^2]^2}{|x-y|^3}dxdy\sim \||u|^2\|_{\dot{H}^{\frac{1}{2}}}^2.
$$
We have
\begin{align}
\|D^{\frac{1}{2}}(|u|^2)\|_{L_t^2L_x^2}^2\leq  C\sup_{t\in[0,T]}\|u(x,t)\|^2_{\dot{H}^{\frac{1}{2}}_x}.
\end{align}
Especially, if $h(|u|^2)=|u|^2$, we have
\begin{align}
\|D^{\frac{1}{2}}(|u|^2)\|_{L_t^2L_x^2}^2+\|D^{\frac{1}{2}}(|u|^3)\|_{L_t^2L_x^2}^2\leq  C\sup_{t\in[0,T]}\|u(x,t)\|^2_{\dot{H}^{\frac{1}{2}}_x}.
\end{align}

\subsection{Interaction Morawetz inequality in dimension $N=1$}
\qquad Besides the assumptions on $h(s)$ and $F(s)$ in (\ref{352}), we assume that
$$
V'(x)\geq 0,\quad W'(x)\geq 0\quad {\rm for \ all}\quad x\in \mathbb{R}.
$$

Let
\begin{align}
a(x):=erf(\frac{x}{\epsilon})=\int_0^{\frac{x}{\epsilon}}e^{-t^2}dt\label{217w1}
\end{align}
and
\begin{align}
M_a(t)=\frac{1}{2}\int_{\mathbb{R}}\int_{\mathbb{R}}a(x-y)|u(y,t)|^2\Im(u(x,t)\bar{u}_x(x,t))dxdy.\label{217w2}
\end{align}
In convenience, we denote
\begin{align}
\rho(x)=\frac{1}{2}|u(x,t)|^2,\quad p(x)=\Im(u(x,t)\bar{u}_x(x,t)).\label{217w3}
\end{align}
After some elementary computations, we obtain

\begin{align}
\frac{d}{dt}M_a(t)&=\int_{\mathbb{R}}\int_{\mathbb{R}}\frac{1}{\epsilon}e^{-\frac{(x-y)^2}{\epsilon^2}}\frac{\rho(y)}{\rho(x)}(\partial_x\rho(x))^2dxdy\nonumber\\
&\quad+\frac{1}{2}\int_{\mathbb{R}}\int_{\mathbb{R}}\frac{1}{\epsilon}e^{-\frac{(x-y)^2}{\epsilon^2}}
\left(\sqrt{\frac{\rho(y)}{\rho(x)}}p(x)-\sqrt{\frac{\rho(x)}{\rho(y)}}p(y)\right)^2dxdy\nonumber\\
&\quad+\int_{\mathbb{R}}\int_{\mathbb{R}}\frac{1}{\epsilon}e^{-\frac{(x-y)^2}{\epsilon^2}}[-\partial_{xx}\rho(x)]\rho(y)dxdy\nonumber\\
&\quad-\int_{\mathbb{R}}\int_{\mathbb{R}}\frac{1}{\epsilon}e^{-\frac{(x-y)^2}{\epsilon^2}}h'(|u(x,t)|^2)|u(x,t)|^2(h(|u(x,t)|^2))_{xx}dx|u(y,t)|^2dy\nonumber\\
&\quad+\frac{1}{2}\int_{\mathbb{R}}\int_{\mathbb{R}}\frac{1}{\epsilon}e^{-\frac{(x-y)^2}{\epsilon^2}}[(h(|u(x,t)|^2))_x]^2dx|u(y,t)|^2dy\nonumber\\
&\quad+\frac{1}{2}\int_{\mathbb{R}}\int_{\mathbb{R}}\frac{1}{\epsilon}e^{-\frac{(x-y)^2}{\epsilon^2}}[G(|u(x,t)|^2)-F(|u(x,t)|^2)|u(x,t)|^2]dx|u(y,t)|^2dy\nonumber\\
&\quad+\frac{1}{2}\int_{\mathbb{R}}\int_{\mathbb{R}}a(x-y)[V_x+\frac{1}{2}(W_x*|u|^2)]|u(x,t)|^2dx|u(y,t)|^2dy.\label{217w4}
\end{align}
Noticing that
\begin{align*}
\int_{\mathbb{R}}\int_{\mathbb{R}}\frac{1}{\epsilon}e^{-\frac{(x-y)^2}{\epsilon^2}}[-\partial_{xx}\rho(x)]\rho(y)dxdy=\int_{\mathbb{R}}\xi^2\hat{\rho}^2(\xi)
e^{-\epsilon\xi^2}d\xi\geq 0,
\end{align*}
$V'(x)\geq 0$ and $W'(x)\geq 0$, from (\ref{217w4}), we can get
\begin{align}
&\quad\int_{\mathbb{R}}\int_{\mathbb{R}}\frac{1}{\epsilon}e^{-\frac{(x-y)^2}{\epsilon^2}}\frac{\rho(y)}{\rho(x)}(\partial_x\rho(x))^2dxdy\nonumber\\
&\quad-\int_{\mathbb{R}}\int_{\mathbb{R}}\frac{1}{\epsilon}e^{-\frac{(x-y)^2}{\epsilon^2}}h'(|u(x,t)|^2)|u(x,t)|^2(h(|u(x,t)|^2))_{xx}dx|u(y,t)|^2dy\nonumber\\
&\quad+\frac{1}{2}\int_{\mathbb{R}}\int_{\mathbb{R}}\frac{1}{\epsilon}e^{-\frac{(x-y)^2}{\epsilon^2}}[(h(|u(x,t)|^2))_x]^2dx|u(y,t)|^2dy\nonumber\\
&\quad+\frac{1}{2}\int_{\mathbb{R}}\int_{\mathbb{R}}\frac{1}{\epsilon}e^{-\frac{(x-y)^2}{\epsilon^2}}[G(|u(x,t)|^2)-F(|u(x,t)|^2)|u(x,t)|^2]dx|u(y,t)|^2dy\nonumber\\
&\leq \frac{d}{dt}M_a(t).\label{217w5}
\end{align}
Letting $\epsilon\rightarrow 0$, and integrating (\ref{217w5}) from $0$ to $T$, we have
\begin{align*}
&\quad\int_0^T\int_{\mathbb{R}}(\partial_x\rho(x))^2dxdt-\int_0^T\int_{\mathbb{R}}|u(x,t)|^4h'(|u(x,t)|^2)(h(|u(x,t)|^2))_{xx}dxdt\nonumber\\
&\quad+\frac{1}{2}\int_0^T\int_{\mathbb{R}}|u(x,t)|^2[(h(|u(x,t)|^2))_x]^2dxdt\nonumber\\
&\quad+\frac{1}{2}\int_0^T\int_{\mathbb{R}}[G(|u(x,t)|^2)-F(|u(x,t)|^2)|u(x,t)|^2]|u(x,t)|^2dxdt\nonumber\\
&=\int_0^T\int_{\mathbb{R}}(\partial_x\rho(x))^2dxdt+\int_0^T\int_{\mathbb{R}}[|u(x,t)|^4h'(|u(x,t)|^2)]_x[h(|u(x,t)|^2)]_xdxdt\nonumber\\
&\quad+\frac{1}{2}\int_0^T\int_{\mathbb{R}}|u(x,t)|^2[(h(|u(x,t)|^2))_x]^2dxdt\nonumber\\
&\quad+\frac{1}{2}\int_0^T\int_{\mathbb{R}}[G(|u(x,t)|^2)-F(|u(x,t)|^2)|u(x,t)|^2]|u(x,t)|^2dxdt\nonumber\\
&\leq \overline{\lim_{\epsilon\rightarrow 0}}\sup_{t\in [0,T]}|M(t)|.
\end{align*}
Therefore,
\begin{align}
&\quad\int_0^T\int_{\mathbb{R}}\left\{[5h'(|u|^2)+2h''(|u|^2)|u|^2]h'(|u|^2)|u|^2+1\right\}[\partial_x|u(x,t)|^2]^2dxdt\nonumber\\
&+\int_0^T\int_{\mathbb{R}}[G(|u(x,t)|^2)-F(|u(x,t)|^2)|u(x,t)|^2]|u(x,t)|^2dxdt\nonumber\\
&\leq C'\sup_{t\in [0,T]}|M(t)|\leq C'\sup_{t\in [0,T]}\|u(x,t)\|^2_{\dot{H}_x^{\frac{1}{2}}}\|u(x,t)\|_{L_x^2}^2\leq
C\sup_{t\in [0,T]}\|u(x,t)\|^2_{\dot{H}_x^{\frac{1}{2}}}.\label{2181}
\end{align}

{\bf Theorem 5.} ({\bf Interaction Morawetz estimates}) {\it  Let $u(x,t)$ be the $H^{\frac{1}{2}}$ solution of (\ref{1}) on the space-time slab $\mathbb{R}^N\times [0,T]$.

1. If $N\geq 3$, under the assumptions of (\ref{352}) and (\ref{353}), then
 \begin{align}
\|D^{-\frac{N-3}{2}}\left(|u(x,t)|^2+\sqrt{|u(x,t)|^2\tilde{H}(|u(x,t)|^2)}\right)\|_{L^2_tL^2_x}^2\leq C\sup_{t\in[0,T]}\|u(x,t)\|_{\dot{H}^{1/2}_x}^2.\label{2226}
\end{align}
Here $\tilde{H}(|u|^2)=\int_0^{|u|^2}[h'(s)]^2sds$.

2. If $N=2$, under the assumptions of (\ref{352}) and (\ref{353}), then
\begin{align}
&\quad\int_0^T\int_{\mathbb{R}^2\otimes\mathbb{R}^2}\frac{[|u(x,t)|^2-|u(y,t)|^2]^2}{|x-y|^3}dxdydt\nonumber\\
&+\int_0^T\int_{\mathbb{R}^2\otimes\mathbb{R}^2}\frac{[|u(x,t)|^2-|u(y,t)|^2][\tilde{H}(|u(x,t)|^2)-\tilde{H}(|u(y,t)|^2)]}{|x-y|^3}dxdydt\nonumber\\
&\leq C\sup_{t\in[0,T]}\|u(x,t)\|^2_{\dot{H}^{\frac{1}{2}}_x}\label{218x7}
\end{align}
Here $\tilde{H}(|u|^2)=\int_0^{|u|^2}[h'(s)]^2sds$.

3. If $N=1$, $G(s)\geq F(s)s$ for $s\geq 0$ and $V'(x)\geq 0$, $W'(x)\geq 0$ for $x\in \mathbb{R}$, then
 \begin{align}
&\quad\int_0^T\int_{\mathbb{R}}\left\{[5h'(|u|^2)+2h''(|u|^2)|u|^2]h'(|u|^2)|u|^2+1\right\}[\partial_x|u(x,t)|^2]^2dxdt\nonumber\\
&+\int_0^T\int_{\mathbb{R}}[G(|u(x,t)|^2)-F(|u(x,t)|^2)|u(x,t)|^2]|u(x,t)|^2dxdt\nonumber\\
&\leq C\sup_{t\in [0,T]}\|u(x,t)\|^2_{\dot{H}_x^{\frac{1}{2}}}\label{2181}
\end{align}
}

{\bf Remark 5.1.} If $h(|u|^2)\equiv 0$, our results meet with those of \cite{Colliander1, Colliander4}. If $h(|u|^2)=|u|^2$, our results meet with those of \cite{Selvitella}. If $h(|u|^2)=|u|^{2\alpha}$, under the assumptions of Theorem 5, we have
 \begin{align*}
\|D^{-\frac{N-3}{2}}\left(|u(x)|^2+|u(x,t)|^{2\alpha+1}\right)\|_{L^2_tL^2_x}^2\leq C\sup_{t\in[0,T]}\|u(x,t)\|_{\dot{H}^{1/2}_x}^2.
\end{align*}

\section{Classic scattering theory for (\ref{semilinear1}) in defocusing case and arbitrary space dimension}
\qquad In this section, applying the results of Theorem 3 and Theorem 4, we will establish scattering theory in $L^2(\mathbb{R}^N)$ and $\Sigma$ ($N\geq 1$) under certain assumptions.

\subsection{Classic scattering theory in $L^2(\mathbb{R}^N)$ for (\ref{semilinear1}) in defocusing case and arbitrary space dimension}
\qquad In this subsection, we will establish classic scattering theory in $L^2(\mathbb{R}^N)$  for (\ref{semilinear1}) in defocusing case and arbitrary space dimension.

{\bf Theorem 6.} {\it Let $u\in C(\mathbb{R},\Sigma)$ be the solution of (\ref{semilinear1}) in defocusing case, i.e., $F(s)\leq 0$ for $s\geq 0$, $V(x)\leq 0$ and $W(x)\leq 0$, $W(x)$ is even for $x\in \mathbb{R}^N$,  $N\geq 1$, and $u_0\in\Sigma$.

Assume that there exist $C>0$, $\theta_1$, $p_1$, $\theta_2$ and $p_2$ such that
\begin{align}
&[|F(s)|s^{\frac{1}{2}}]^{\theta_1}\leq Cs,\ [|F(s)|s^{\frac{1}{2}}]^{p_1}\leq C|G(s)|,\quad 0<s<1,\label{713x2}\\
&[|F(s)|s^{\frac{1}{2}}]^{\theta_2}\leq Cs,\ [|F(s)|s^{\frac{1}{2}}]^{p_2}\leq C|G(s)|,\quad s>1,\label{6252}
\end{align}
and there exist $c_1$, $c_2$,  $V_1(x)$, $V_2(x)$, $W_1(x)$ and $W_2(x)$ such that
\begin{align}
&V(x)=V_1(x)+V_2(x),\quad c_1(|V_1(x)|+|V_2(x)|)\leq |V(x)|,\label{622x3}\\
&W(x)=W_1(x)+W_2(x),\quad c_2(|W_1(x)|+|W_2(x)|)\leq |W(x)|.\label{622x3''}
\end{align}

In addition, suppose that there exist admissible pairs $(q_1,r_1)$, $(q_2,r_2)$, $(q_3,r_3)$, $(\tilde{q}_1, \tilde{r}_1)$ and $(\tilde{q}_2, \tilde{r}_2)$
such that
\begin{align}
&V_1(x)\in L^{\frac{r_1}{r_1-2}}(\mathbb{R}^N),\quad V_2(x)\in  L^{\frac{r_2}{r_2-2}}(\mathbb{R}^N),\label{7131}\\
&W_1(x)\in L^{\frac{\tilde{r}_1}{2(\tilde{r}_1-2)}}(\mathbb{R}^N),\quad W_2(x)\in  L^{\frac{\tilde{r}_2}{2(\tilde{r}_2-2)}}(\mathbb{R}^N),\label{622w1}
\end{align}
and
\begin{align}
q'_1>1,\quad q'_2>1,\quad \frac{2q'_3(r'_3-\theta_1)}{r'_3(p_1-\theta_1)}>1,\quad \frac{2q'_3(r'_3-\theta_2)}{r'_3(p_2-\theta_2)}>1,\quad
\tilde{q}'_1>1,\quad \tilde{q}'_2>1\label{8123}
\end{align}
if $[(N+2)G(s)-NF(s)s]\geq 0$ for $s\geq 0$, $[2V+(x\cdot \nabla V)]\geq 0$ and $[2W+(x\cdot \nabla W)]\geq 0$ for $x\in \mathbb{R}^N$, while
\begin{align}
&\frac{(2-l)q'_1}{2}>1,\quad \frac{(2-l)q'_2}{2}>1,\quad \frac{(2-l)q'_3(r'_3-\theta_1)}{r'_3(p_1-\theta_1)}>1,\label{8121}\\ &\frac{(2-l)q'_3(r'_3-\theta_2)}{r'_3(p_2-\theta_2)}>1,\quad \frac{(2-l)\tilde{q}'_1}{2}>1,\quad \frac{(2-l)\tilde{q}'_2}{2}>1\label{8122}
\end{align}
if at least one of the following cases holds:

(i) $-k_1|G(s)|\leq (N+2)G(s)-NF(s)s\leq 0$ for some $k_1>0$;

(iv) $-k_2|V|\leq 2V+(x\cdot \nabla V)\leq 0$ for some $k_2>0$;

(v) $-k_3|W|\leq 2W+(x\cdot \nabla W)\leq 0$ for some $k_3>0$.

Here
\begin{align}
l=\max(k_1, k_2, k_3),\label{311xj2}
\end{align}
$q'_j$, $r'_j$, $\tilde{q}'_m$, $\tilde{r}'_m$ are the conjugated exponents of $q_j$, $r_j$, $\tilde{q}_m$, $\tilde{r}_m$ respectively.

Then there exists $u_+\in L^2(\mathbb{R}^N)$ such that
$$e^{it\Delta}u(t)\longrightarrow u_+\quad {\rm in}\quad L^2(\mathbb{R}^N)\quad {\rm as}\quad t\rightarrow +\infty.$$
}

{\bf Proof:} Duhamel's principle implies that
$$
u(t)=e^{-it\Delta}u_0-i\int_0^te^{-i(t-s)\Delta}\left(V(x)u(s)+F(|u|^2)u(s)+(W*|u|^2)u(s)\right)ds.
$$

By Strichartz estimates, using H\"{o}lder inequality, for any $0<t<\tau$, we obtain
\begin{align}
&\quad\|e^{it\Delta}u(t)-e^{i\tau\Delta}u(\tau)\|_{L^2}\nonumber\\
&\leq \|\int_t^{\tau} e^{is\Delta}V(x)u(s)ds\|_{L^2}+\|\int_t^{\tau} e^{is\Delta}F(|u|^2)u(s)ds\|_{L^2}+\|\int_t^{\tau} e^{is\Delta}(W*|u|^2)u(s)ds\|_{L^2}\nonumber\\
&\leq \sum_{j=1}^2\|\int_t^{\tau} e^{is\Delta}V_j(x)u(s)ds\|_{L^2}+\|\int_t^{\tau} e^{is\Delta}F(|u|^2)u(s)ds\|_{L^2}\nonumber\\
&\quad+\sum_{m=1}^2\|\int_t^{\tau} e^{is\Delta}(W_m*|u|^2)u(s)ds\|_{L^2}\nonumber\\
&\leq C\sum_{j=1}^2\left(\int_t^{\tau}\left(\int_{\mathbb{R}^N}|V_j(x)u|^{r'_j}dx\right)^{\frac{q'_j}{r'_j}}dt\right)^{\frac{1}{q'_j}}+
C\left(\int_t^{\tau}\left(\int_{\mathbb{R}^N}[|F(|u|^2)||u|]^{r'_3}dx\right)^{\frac{q'_3}{r'_3}}dt\right)^{\frac{1}{q'_3}}\nonumber\\
&\quad+C\sum_{m=1}^2\left(\int_t^{\tau}\left(\int_{\mathbb{R}^N}[(|W_m|*|u|^2)|u|]^{\tilde{r}'_m}dx
\right)^{\frac{\tilde{q}'_m}{\tilde{r}'_m}}dt\right)^{\frac{1}{\tilde{q}'_m}}\nonumber\\
&:=(I)+(II)+(III).\label{72j1}
\end{align}

Using H\"{o}lder inequality, it is easy to get
\begin{align}
(I)&\leq C\left(\int_t^{\tau}\left(\int_{\mathbb{R}^N}|V_1(x)||u|^2dx\right)^{\frac{q'_1}{2}}
\left(\int_{\mathbb{R}^N}|V_1(x)|^{\frac{r'_1}{2-r'_1}}dx\right)^{\frac{q'_1(2-r'_1)}{2r'_1}}dt\right)^{\frac{1}{q'_1}}\nonumber\\
&\quad+C\left(\int_t^{\tau}\left(\int_{\mathbb{R}^N}|V_2(x)||u|^2dx\right)^{\frac{q'_2}{2}}
\left(\int_{\mathbb{R}^N}|V_2(x)|^{\frac{r'_2}{2-r'_2}}dx\right)^{\frac{q'_2(2-r'_2)}{2r'_2}}dt\right)^{\frac{1}{q'_2}}\nonumber\\
&\quad \longrightarrow 0\quad {\rm as}\quad  t,\ \tau \rightarrow +\infty,\label{7132}\\
(II)&\leq
C\left(\int_t^{\tau}\left(\int_{\{|u|\leq 1\}}[|F(|u|^2)||u|]^{\theta_1}dx\right)^{\frac{q'_3}{\tau'_1r'_3}}
\left(\int_{\{|u|\leq 1\}}[|F(|u|^2)||u|]^{p_1}dx\right)^{\frac{q'_3}{\tau_1r'_3}}dt\right)^{\frac{1}{q'_3}}\nonumber\\
&\quad+
C\left(\int_t^{\tau}\left(\int_{\{|u|>1\}}[|F(|u|^2)||u|]^{\theta_2}dx\right)^{\frac{q'_3}{\tau'_2r'_3}}
\left(\int_{\{|u|>1\}}[|F(|u|^2)||u|]^{p_2}dx\right)^{\frac{q'_3}{\tau_2r'_3}}dt\right)^{\frac{1}{q'_3}}\nonumber\displaybreak\\
&\leq
C\left(\int_t^{\tau}\left(\int_{\{|u|\leq 1\}}|u|^2dx\right)^{\frac{q'_3}{\tau'_1r'_3}}\left(\int_{\{|u|\leq 1\}}|G(|u|^2)|dx\right)^{\frac{q'_3}{\tau_1r'_3}}
dt\right)^{\frac{1}{q'_3}}\nonumber\\
&\quad+ C\left(\int_t^{\tau}\left(\int_{\{|u|>1\}}|u|^2dx\right)^{\frac{q'_3}{\tau'_2r'_3}}\left(\int_{\{|u|>1\}}|G(|u|^2)|dx\right)^{\frac{q'_3}{\tau_2r'_3}}
dt\right)^{\frac{1}{q'_3}}\nonumber\\
&\leq
C\left(\int_t^{\tau}\left(\int_{\mathbb{R}^N}G(|u|^2)dx\right)^{\frac{q'_3}{\tau_1r'_3}}
dt\right)^{\frac{1}{q'_3}}+C\left(\int_t^{\tau}\left(\int_{\mathbb{R}^N}|G(|u|^2)|dx\right)^{\frac{q'_3}{\tau_2r'_3}}
dt\right)^{\frac{1}{q'_3}}\nonumber\\
&\quad \longrightarrow 0\quad {\rm as}\quad t,\ \tau\rightarrow +\infty,\label{713x3}\\
(III)&\leq \sum_{m=1}^2C\left\{\int_t^{\tau}\left(\int_{\mathbb{R}^N}(|W_m|*|u|^2)|u|^2dx\right)^{\frac{\tilde{q}'_m}{2}} \left(\int_{\mathbb{R}^N}|u(x)|^{\frac{r'_m}{r'_m-1}}dx\right)^{\frac{\tilde{q}'_m(\tilde{r}'_m-1)}{\tilde{r}'_m}}\right.\nonumber\\
&\qquad\qquad\qquad\left. \left(\int_{\mathbb{R}^N}\int_{\mathbb{R}^N}|W_m(x-y)|^{\frac{\tilde{r}'_m}{2-\tilde{r}'_m}}dydx\right)^{\frac{\tilde{q}'_m(2-\tilde{r}'_m)}{2\tilde{r}'_m}}
dt\right\}^{\frac{1}{\tilde{q}'_m}}\nonumber\\
&\quad \longrightarrow 0\quad {\rm as}\quad t,\  \tau\rightarrow +\infty,\label{711}
\end{align}
because
$$
\int_{\mathbb{R}^N}|u|^2dx=\int_{\mathbb{R}^N}|u_0|^2dx,\quad \int_{\mathbb{R}^N}|\nabla u|^2dx\leq C,\quad
\int_{\mathbb{R}^N}|u(x)|^{\frac{\tilde{r}'_m}{\tilde{r}'_m-1}}dx\leq C
$$
by the results of Section 3 and Section 4, moreover,
$$
\int_{\mathbb{R}^N}|V(x)||u|^2dx\leq \frac{C}{t^2},\quad \int_{\mathbb{R}^N}|G(|u|^2)|dx\leq \frac{C}{t^2},\quad
\int_{\mathbb{R}^N}(|W|*|u|^2)|u|^2dx\leq \frac{C}{t^2}
$$
and (\ref{8123}) in Case 1, while
$$
\int_{\mathbb{R}^N}|V(x)||u|^2dx\leq \frac{C}{t^{2-l}},\quad \int_{\mathbb{R}^N}|G(|u|^2)|dx\leq \frac{C}{t^{2-l}},\quad
\int_{\mathbb{R}^N}(|W|*|u|^2)|u|^2dx\leq \frac{C}{t^{2-l}}
$$
and (\ref{8121}), (\ref{8122}) in Case 2. Here
$$
\frac{1}{\tau_1}=\frac{r'_3-\theta_1}{p_1-\theta_1},\quad \frac{1}{\tau'_1}=\frac{p_1-r'_3}{p_1-\theta_1},\quad
 \frac{1}{\tau_2}=\frac{r'_3-\theta_2}{p_2-\theta_2},\quad \frac{1}{\tau'_2}=\frac{p_2-r'_3}{p_2-\theta_2}.
$$

Consequently, there exists $u_+\in L^2(\mathbb{R}^N)$ such that $$\|e^{it\Delta} u(t)-u_+\|_{L^2}\rightarrow 0\quad {\rm as}\quad t\rightarrow +\infty.$$
That is, every solution in $\Sigma$ of (\ref{semilinear1}) has scattering state in $L^2(\mathbb{R}^N)$.
\hfill $\Box$

{\bf Remark 6.1.} 1. A special case in the assumptions of Theorem 6 is $\theta_1=\theta_2=\theta$, $p_1=p_2=p$. For example, if $F(|u|^2)u=b|u|^{2\beta}u$, then $\theta_1=\theta_2=\frac{2}{2\beta+1}$, $p_1=p_2=\frac{2\beta+2}{2\beta+1}$, and the assumptions of Theorem 6 can be satisfied.

 2. In the proof of Theorem 6, we take different admissible pairs in Strichartz estimates for different terms on the right of Duhamel's formula in order to keep the terms containing $V(x)u$, $F(|u|^2)u$ and $(W*|u|^2)u$ independent each other. Consequently, Theorem 6 can deduce scattering theory in $L^2(\mathbb{R}^N)$ for Cauchy problem of the equation contains one of $V(x)u$, $F(|u|^2)u$ and $(W*|u|^2)u$ directly.

{\bf Corollary 6.1.} {\it Let $u$ be the solution of the following problem
\begin{equation}
\label{812x1} \left\{
\begin{array}{lll}
iu_t=\Delta u+V(x)u,\ x\in \mathbb{R}^N,\ t>0\\
u(x,0)=u_0(x)\in \Sigma,\quad x\in \mathbb{R}^N.
\end{array}\right.
\end{equation}
Assume that $V(x)\leq 0$ for $x\in \mathbb{R}^N$, $N\geq 1$, and (\ref{622x3}), (\ref{7131}), (\ref{8123}) and (\ref{8121}) hold. Then there exists $u_+\in L^2(\mathbb{R}^N)$ such that
$$\|e^{it\Delta} u(t)-u_+\|_{L^2}\rightarrow 0\quad {\rm as}\quad t\rightarrow +\infty.$$}

{\bf Corollary 6.2.} {\it Let $u$ be the solution of the following problem
\begin{equation}
\label{812x2} \left\{
\begin{array}{lll}
iu_t=\Delta u+F(|u|^2)u,\ x\in \mathbb{R}^N,\ t>0\\
u(x,0)=u_0(x)\in \Sigma,\quad x\in \mathbb{R}^N.
\end{array}\right.
\end{equation}
Assume that $F(s)$ satisfies (G), $F(s)\leq 0$ for $s\geq 0$, $N\geq 1$,  and (\ref{713x2}), (\ref{6252}), (\ref{8123}), (\ref{8121}) and (\ref{8122}) hold. Then there exists $u_+\in L^2(\mathbb{R}^N)$ such that
$$\|e^{it\Delta} u(t)-u_+\|_{L^2}\rightarrow 0\quad {\rm as}\quad t\rightarrow +\infty.$$}

{\bf Corollary 6.3.} {\it Let $u$ be the solution of the following problem
\begin{equation}
\label{812x3} \left\{
\begin{array}{lll}
iu_t=\Delta u+(W*|u|^2)u,\ x\in \mathbb{R}^N,\ t>0\\
u(x,0)=u_0(x)\in \Sigma,\quad x\in \mathbb{R}^N.
\end{array}\right.
\end{equation}
Assume that $W(x)$ is even and $W(x)\leq 0$ for $x\in \mathbb{R}^N$, $N\geq 1$, and (\ref{622x3''}), (\ref{622w1}), (\ref{8123}) and (\ref{8122}) hold. Then there exists $u_+\in L^2(\mathbb{R}^N)$ such that
$$\|e^{it\Delta} u(t)-u_+\|_{L^2}\rightarrow 0\quad {\rm as}\quad t\rightarrow +\infty.$$}

2. If the nonlinearities of a semilinear Sch\"{o}dinger equation are combined by any two terms of $V(x)u$, $F(|u|^2)u$ and $(W*|u|^2)u$, then we also can establish the scattering theory in $L^2(\mathbb{R}^N)$ directly. For example, we have

{\bf Corollary 6.4.} {\it Let $u$ be the solution of
\begin{equation}
\label{812x4} \left\{
\begin{array}{lll}
iu_t=\Delta u+V(x)u+(W*|u|^2)u,\ x\in \mathbb{R}^N,\ t>0\\
u(x,0)=u_0(x)\in \Sigma,\quad x\in \mathbb{R}^N.
\end{array}\right.
\end{equation}
Assume that $V(x)\leq 0$ and $W(x)\leq 0$ for $x\in \mathbb{R}^N$, $N\geq 1$, $W(x)$ is even, and (\ref{622x3})--(\ref{8122}) hold. Then there exists $u_+\in L^2(\mathbb{R}^N)$ such that
$$\|e^{it\Delta} u(t)-u_+\|_{L^2}\rightarrow 0\quad {\rm as}\quad t\rightarrow +\infty.$$}

As a corollary of Theorem 6, we give the scattering theory in $L^2(\mathbb{R}^N)$ of (\ref{224x2}) below.

{\bf Corollary 6.5.} {\it Assume that $u(x,t)$ is the solution of (\ref{224x2}) and $u_0\in \Sigma$.
Then there exists $u_+\in L^2(\mathbb{R}^N)$ such that
$$\|e^{it\Delta} u(t)-u_+\|_{L^2}\rightarrow 0\quad {\rm as}\quad t\rightarrow +\infty$$

if one of the following cases holds:

(I). $N\geq 2$, $\frac{4}{3}<m<2$, $m<n<4$, $\frac{4}{3}<m<N\beta<2^*$, $8<4m+n$;

(II). $N\geq 2$, $\beta_0<N\beta<m<2$, $\beta_0<N\beta<n<4$, $4<2N\beta+m$, $8<4N\beta+n$;

(III). $N\geq 2$, $\frac{8}{5}<n<m<2$, $n<N\beta<2^*$.

Here
$$
\beta_0=\frac{4-3N+\sqrt{9N^2+40N+16}}{8N},
$$
$2^*=\frac{2N}{N-2}$ if $N\geq 3$ and $2^*=+\infty$ if $N=2$.

}

{\bf Proof:} Let
\begin{align*}
&V_1(x)=\left\{
\begin{array}{lll}
-\frac{1}{|x|^m},\quad 0<|x|\leq 1,\\
0,\quad |x|>1,
\end{array}\right.\quad {\rm and}\quad
V_2(x)=\left\{
\begin{array}{lll}
0,\quad 0<|x|\leq 1,\\
-\frac{1}{|x|^m},\quad |x|>1,
\end{array}\right.\\
&F(|u|^2)u=|u|^{2\beta}u,\quad \theta_1=\theta_2=\frac{2}{2\beta+1},\quad p_1=p_2=\frac{2\beta+2}{2\beta+1},\\
&W_1(x)=\left\{
\begin{array}{lll}
-\frac{1}{|x|^n},\quad 0<|x|\leq 1,\\
0,\quad |x|>1,
\end{array}\right.\quad {\rm and}\quad
W_2(x)=\left\{
\begin{array}{lll}
0,\quad 0<|x|\leq 1,\\
-\frac{1}{|x|^n},\quad |x|>1.
\end{array}\right.
\end{align*}

Since \begin{align*}
&-(2-m)|V(x)|=2V(x)+x\cdot \nabla V(x)=-\frac{(2-m)}{|x|^m}<0,\nonumber\\
&2W(x)+x\cdot \nabla W(x)=\frac{(n-2)}{|x|^n}=(n-2)|W(x)|,\\
&NF(s)s-(N+2)G(s)=-\frac{(2-N\beta)}{\beta+1}|u|^{2\beta+2}=-(2-N\beta)|G(s)|,
\end{align*}
it belongs to Case 2 of Theorem 6.

We can take $r'_1$, $r'_2$, $r'_3$, $\tilde{r}'_1$ and $\tilde{r}'_2$ respectively as follows:

(I).  $\min(m, N\beta,n)=m$.
\begin{align*}
&\frac{2N}{N+2}<r'_1<\frac{2N}{N+m},\quad \frac{2N}{N+2}<\tilde{r}'_1<\frac{4N}{2N+n},\\
&\frac{2N}{N+m}<r'_2<\frac{2N}{N+4-2m},\quad \frac{4N}{2N+n}<\tilde{r}'_2<\frac{2N}{N+4-2m},\\
&\max(\frac{2N}{N+2}, \frac{4m-2N\beta}{2m+4m\beta-4\beta-N\beta})<r'_3<\frac{2\beta+2}{2\beta+1}\quad {\rm if}\quad 2m>N\beta,\\
& \frac{2N}{N+2}<r'_3<\frac{2\beta+2}{2\beta+1}\quad {\rm if}\quad 2m\leq N\beta,\quad 2m+4m\beta-4\beta-N\beta\geq 0,\\
&\frac{2N}{N+2}<r'_3<\min(\frac{2\beta+2}{2\beta+1},\frac{2N\beta-4m}{4\beta+N\beta-2m-4m\beta})\\
&\qquad{\rm if}\ 2m<N\beta,\ 2m+4m\beta-4\beta-N\beta< 0;
\end{align*}

(II). $\min(m, N\beta,n)=N\beta$.
\begin{align*}
&\frac{2N}{N+2}<r'_1<\frac{2N}{N+m},\quad \frac{2N}{N+2}<\tilde{r}'_1<\frac{4N}{2N+n},\\
&\frac{2N}{N+m}<r'_2<\frac{2N}{N+4-2N\beta},\quad \frac{4N}{2N+n}<\tilde{r}'_2<\frac{2N}{N+4-2N\beta},\\
&\max(\frac{2N}{N+2},\frac{2}{2\beta+1}, \frac{2N}{N+4N\beta-4})<r'_3<\frac{2\beta+2}{2\beta+1};
\end{align*}

(III). $\min(m, N\beta,n)=n$.
\begin{align*}
&\frac{2N}{N+2}<r'_1<\frac{2N}{N+m},\quad \frac{2N}{N+2}<\tilde{r}'_1<\frac{4N}{2N+n},\\
&\frac{2N}{N+m}<r'_2<\frac{2N}{N+4-2n},\quad \frac{4N}{2N+n}<\tilde{r}'_2<\frac{2N}{N+4-2n},\\
& \max(\frac{2N}{N+2},\frac{4n-2N\beta}{2n+4n\beta-4\beta-N\beta})<r'_3<\frac{2\beta+2}{2\beta+1}\quad {\rm if}\quad 2n>N\beta,\\
& \frac{2N}{N+2}<r'_3<\frac{2\beta+2}{2\beta+1}\quad {\rm if}\quad 2n\leq N\beta,\quad 2n+4n\beta-4\beta-N\beta\geq 0,\\
&\frac{2N}{N+2}<r'_3<\min(\frac{2\beta+2}{2\beta+1},\frac{2N\beta-4n}{4\beta+N\beta-2n-4n\beta})\\
& \qquad {\rm if}\ 2n<N\beta,\ 2n+4n\beta-4\beta-N\beta< 0;
\end{align*}

It is easy to verify the assumptions of Theorem 6 and establish scattering theory in $L^2(\mathbb{R}^N)$ for (\ref{224x2}).\hfill $\Box$

{\bf Remark 6.2.}  Our idea can be applied to deal with the following problem:
\begin{equation*}
\left\{
\begin{array}{lll}
iu_t=\Delta u+\sum_{m=1}^MV_m(x)u+\sum_{k=1}^KF_k(|u|^2)u+\sum_{l=1}^L(W_l*|u|^2)u,\ x\in \mathbb{R}^N,\ t>0\\
u(x,0)=u_0(x),\quad x\in \mathbb{R}^N.
\end{array}\right.
\end{equation*}
And we can obtain the general scattering results similar to Theorem 6.

\subsection{Classic scattering theory in $\Sigma$ for (\ref{semilinear1}) in defocusing case and arbitrary space dimension}
\qquad In the last part of this paper, we will establish classic scattering theory in $\Sigma$ for the solution of (\ref{semilinear1}) in defocusing case
and arbitrary space dimension.

{\bf Theorem 7.} {\it Let $u\in C(\mathbb{R},\Sigma)$ be the solution of (\ref{semilinear1}) in defocusing case with $u_0\in\Sigma$.
Assume that $V(x)\equiv 0$ and $W(x)\equiv 0$ for $x\in \mathbb{R}^N$, $F(s)$ satisfies (G):

 $({\bf G})$\quad $\frac{|G(s)|}{[s^{\frac{1}{2}}+h(s)]^{2^*}}\rightarrow 0$ as $s\rightarrow +\infty$, where $G(s)=\int_0^sF(\eta)d\eta$,\\
and there exist $C>0$, $\theta_1$, $p_1$, $\theta_2$, $p_2$, $2<r<\frac{2N}{N-2}$ if $N\geq 3$,
$2<r<+\infty$ if $N=1, 2$, $0<l<2$, such that
\begin{align}
&\theta_1<\frac{r}{r-2}<p_1,\quad \theta_2<\frac{r}{r-2}<p_2,\label{84x1}\\
&[|F(s)|+|F'(s)|s^{\frac{1}{2}}]^{\theta_1}\leq Cs,\ [|F(s)|+|F'(s)|s^{\frac{1}{2}}]^{p_1}\leq C|G(s)|,\quad 0<s<1,\label{7221}\\
&[|F(s)|+|F'(s)|s^{\frac{1}{2}}]^{\theta_2}\leq Cs,\ [|F(s)|+|F'(s)|s^{\frac{1}{2}}]^{p_2}\leq C|G(s)|,\quad s>1.\label{7222}
\end{align}

Moreover,
\begin{align}
&\frac{4[r(1-\theta_j)+2\theta_j]}{[2N-(N-2)r](p_j-\theta_j)}>1,\quad j=1,2,\label{722x3}
\end{align}
in Case 1: $[NF(s)s-(N+2)G(s)]\leq 0$ for $s\geq 0$,
\begin{align}
&\frac{2(2-l)[r(1-\theta_j)+2\theta_j]}{[2N-(N-2)r](p_j-\theta_j)}>1,\quad j=1,2,\label{722x4}
\end{align}
in Case 2: $0\leq (N+2)G(s)-NF(s)s\leq l|G(s)|$.

Then there exists $u_+\in \Sigma$ such that
$$e^{it\Delta}u(t)\rightarrow u_+\quad {\rm in}\quad \Sigma \quad {\rm as}\quad t\rightarrow +\infty$$
}

 {\bf Proof:} We only prove it in Case (B). The proof in Case (A) can be obtained similarly.

 Let $(q,r)$ be the admissible pair satisfying
 $$
 \frac{2}{q}=N(\frac{1}{2}-\frac{1}{r}),
 $$
 where $2<r<\frac{2N}{N-2}$ if $N\geq 3$, $2<r<+\infty$ if $N=1, 2$.

 First, we prove that
 \begin{align}
\|u\|_{L^q((0,t), W^{1,r})}\leq C \quad {\rm for }\quad t>0. \label{7161}
 \end{align}
Duhamel's principle implies that
$$
u(t)=e^{-it\Delta}u_0-i\int_0^te^{-i(t-s)\Delta}F(|u|^2)u(s)ds.
$$
By Strichartz estimates, using H\"{o}lder's inequality, we have
\begin{align}
&\quad \|u\|_{L^q((0,t), W^{1,r})}\leq C\|u_0\|_{H^1}+C\|F(|u|^2)u\|_{L^{q'}((0,t),W^{1,r'})}\nonumber\\
&\leq C+C\left(\int_0^T\left(\int_{\mathbb{R}^N}[|F(|u|^2)|+|F'(|u|^2)||u|^2]^{\frac{r}{r-2}}dx\right)^{\frac{q(r-2)}{r(q-2)}}dt\right)^{\frac{q-2}{q}}\|u\|_{L^q((0,T), W^{1,r})}\nonumber\\
&\quad+C\left(\int_T^t\left(\int_{\mathbb{R}^N}[|F(|u|^2)|+|F'(|u|^2)||u|^2]^{\frac{r}{r-2}}dx\right)^{\frac{q(r-2)}{r(q-2)}}dt\right)^{\frac{q-2}{q}}
\|u\|_{L^q((T,t), W^{1,r})}\nonumber\\
&\leq C'+C\sum_{j=1}^2\left(\int_T^t\left\{\left(\int_{\mathbb{R}^N}|u|^2dx\right)^{\frac{1}{\tau'_j}}
\left(\int_{\mathbb{R}^N}|G(|u|^2)|dx\right)^{\frac{1}{\tau_j}}\right\}^{\frac{q(r-2)}{r(q-2)}}dt\right)^{\frac{q-2}{q}}
\|u\|_{L^q((T,t), W^{1,r})}\nonumber\\
&\leq C'+C\sum_{j=1}^2\left(\int_T^t
\left(\int_{\mathbb{R}^N}|G(|u(x)|^2)|dx\right)^{\frac{q[r(1-\theta_j)+2\theta_j]}{r(q-2)(p_j-\theta_j)}}dt\right)^{\frac{q-2}{q}}
\|u\|_{L^q((T,t), W^{1,r})}\nonumber\\
&\leq C+\frac{1}{2}\|u\|_{L^q((0,t), W^{1,r})}\quad {\rm if}\quad T \ {\rm is \quad large\quad enough} \label{7162}
\end{align}
because
$$
\int_{\mathbb{R}^N}|G(|u|^2)|dx\leq \frac{C}{t^2},\quad \frac{4[r(1-\theta_j)+2\theta_j]}{[2N-(N-2)r](p_j-\theta_j)}>1,\quad j=1,2,$$
in Case 1, while
$$
\int_{\mathbb{R}^N}|G(|u|^2)|dx\leq \frac{C}{t^{2-l}},\quad \frac{2(2-l)[r(1-\theta_j)+2\theta_j]}{[2N-(N-2)r](p_j-\theta_j)}>1,\quad j=1,2,
$$
in Case 2. Here
$$
\frac{1}{\tau_j}=\frac{r(1-\theta_j)+2\theta_j}{(r-2)(p_j-\theta_j)},\quad \frac{1}{\tau'_j}=\frac{(r-2)(p_j-\theta_j)-[r(1-\theta_j)+2\theta_j]}{(r-2)(p_j-\theta_j)},\quad j=1, 2.
$$
(\ref{7162}) implies (\ref{7161}).

As a byproduct of (\ref{7162}), we get
\begin{align}
&\|F(|u|^2)u\|_{L^{q'}((t,\tau),W^{1,r'})}\longrightarrow 0 \quad {\rm as}\quad t,\ \tau\rightarrow +\infty.\label{891}
\end{align}
Consequently, we obtain
\begin{align}
&\quad \|e^{it\Delta}u(t)-e^{i\tau\Delta}u(\tau)\|_{H^1}\leq \|\int_t^{\tau} e^{is\Delta}F(|u|^2)u(s)ds\|_{H^1}\nonumber\\
&\leq C\|F(|u|^2)u\|_{L^{q'}((t,\tau), W^{1,r'})}\longrightarrow 0 \quad {\rm as}\quad t,\ \tau\rightarrow +\infty\label{7164}
\end{align}
by the result of (\ref{891}).

Therefore, there exists $u_+\in H^1(\mathbb{R}^N)$ such that
$$e^{it\Delta}u(t)\rightarrow u_+\quad {\rm in}\quad H^1(\mathbb{R}^N) \quad {\rm as}\quad t\rightarrow +\infty.$$

Now we will prove that
 \begin{align}
\|(x-2it\nabla) u\|_{L^q((0,t), W^{1,r})}\leq C \quad {\rm  for }\quad t>0. \label{7165}
 \end{align}
Since
$$
(x-2it\nabla) u(t)=e^{-it\Delta} xu_0-i\int_0^te^{-i(t-s)\Delta}(x-2is\nabla)[F(|u|^2)u(s)]ds,
$$
by Strichartz estimates, we obtain
\begin{align}
\|(x-2it\nabla) u\|_{L^q((0,t), L^r)}&\leq C\|xu_0\|_{L^2}+C\|(x-2it\nabla)[F(|u|^2)u]\|_{L^{q'}((0,t),L^{r'})}.\label{7166}
\end{align}

Letting $H(t):=(x-2it\nabla)$, it is easy to verify that
\begin{align*}
&H(t)[F(|u|^2)u]=\partial_u[F(|u|^2)u]H(t)u-\partial_{\bar{u}}[F(|u|^2)u]\overline{H(t)u}
\end{align*}
and
\begin{align}
&\quad\|(x-2it\nabla)[F(|u|^2)u]\|_{L^{q'}((0,t),L^{r'})}\nonumber\\
&\leq
\|\partial_u[F(|u|^2)u]H(t)u\|_{L^{q'}((0,t),L^{r'})}+\|\partial_{\bar{u}}[F(|u|^2)u]\overline{H(t)u}\|_{L^{q'}((0,t),L^{r'})}\nonumber\\
&\leq C\|[|F(|u|^2)+|F'(|u|^2)|u|^2](x-2it\nabla)u\|_{L^{q'}((0,t),L^{r'})}.\label{922}
\end{align}

By (\ref{7166}) and (\ref{922}), we get
\begin{align}
&\quad \|(x-2it\nabla) u\|_{L^q((0,t), L^r)}\nonumber\\
&\leq C\|xu_0\|_{L^2}+C\|[|F(|u|^2)|+|F'(|u|^2)||u|^2]|(x-2is\nabla)u|\|_{L^{q'}((0,t),L^{r'})}\nonumber\\
&\leq C'+C\|[|F(|u|^2)|+|F'(|u|^2)||u|^2]|(x-2is\nabla)u|\|_{L^{q'}((0,t),L^{r'})}.\label{921}
\end{align}

Similar to the discussion of (\ref{7162}) and (\ref{891}), we have (\ref{7165}) and
\begin{align}
 \|[F(|u|^2)+|F'(|u|^2)||u|^2](x-2is\nabla)u\|_{L^{q'}((t,\tau),L^{r'})}\longrightarrow 0\quad {\rm as}\quad t,\ \tau\rightarrow +\infty.\label{7167}
\end{align}
Consequently,
\begin{align}
&\quad\|xe^{it\Delta}u(t)-xe^{i\tau\Delta}u(\tau)\|_{L^2}=\|\int_t^{\tau}e^{is\Delta}(x-2is\nabla)[F(|u|^2)u(s)\nonumber\\
&\leq C\|[|F(|u|^2)+|F'(|u|^2)|u|^2]|(x-2is\nabla)u|\|_{L^{q'}((t,\tau),L^{r'})}\longrightarrow 0\label{7168}
\end{align}
 as $t,\ \tau\rightarrow +\infty$ by the result of  (\ref{7167}).

Hence, there exists $u_+\in \Sigma$ such that
$$e^{it\Delta}u(t)\rightarrow u_+\quad {\rm in}\quad \Sigma \quad {\rm as}\quad t\rightarrow +\infty.$$
That is, if $V(x)\equiv 0$ and $W(x)\equiv 0$, under the assumptions on $F(s)$, every solution with initial data $u_0\in\Sigma$ of (\ref{semilinear1}) has scattering state in $\Sigma$.\hfill $\Box$

{\bf Remark 6.3.}  We would like to give an example to illustrate the results of Theorem 7.
 $$F(|u|^2)u=a_1|u|^{2\beta_1}u+...+a_m|u|^{2\beta_m}u,\quad
 \frac{2-N+\sqrt{N^2+12N+4}}{4N}<\beta_1<...<\beta_m<\frac{2^*}{N},$$
$a_j<0$, $j=1,2,...,m$. Here $2^*=\frac{2N}{N-2}$ if $N\geq 3$, $2^*=+\infty$ if $N=1,2$. Taking
$$\theta_1=\frac{1}{\beta_1},\quad p_1=\frac{\beta_1+1}{\beta_1},\quad \theta_2=\frac{1}{\beta_m},\quad p_2=\frac{\beta_m+1}{\beta_m},$$
and the assumptions of Theorem 7 can be satisfied,  we can obtain the corresponding scattering results.

{\bf Remark 6.4.}  In \cite{Song2}, we considered the following Cauchy problem
\begin{equation}
\label{92w1} \left\{
\begin{array}{lll}
iu_t=\Delta u+F(|u|^2)u-A|u|^{2^*-2}u,\ x\in \mathbb{R}^N,\ t>0\\
u(x,0)=u_0(x)\in \Sigma,\quad x\in \mathbb{R}^N,
\end{array}\right.
\end{equation}
and established the following theorem:

{\bf Theorem 2 of \cite{Song2}.(Scattering theory in $\Sigma$)} {\it Let $u\in C(\mathbb{R},\Sigma)$ be the global solution of (\ref{92w1}), $N\geq 3$, $A>0$ and $u_0\in\Sigma$.
Suppose that $NF(s)s \leq (N+2)G(s)\leq 0$ for $s\geq 0$, and there exist $C>0$, $\theta_1<\frac{N}{2}<p_1$ and $\theta_2<\frac{N}{2}<p_2$ such that
\begin{align}
&[|F(s)|+|F'(s)|s]^{\theta_1}\leq Cs,\ [|F(s)|+|F'(s)|s]^{p_1}\leq C|G(s)|,\quad 0<s<1,\label{92w2}\\
&[|F(s)|+|F'(s)|s]^{\theta_2}\leq Cs,\ [|F(s)|+|F'(s)|s]^{p_2}\leq C|G(s)|,\quad s>1.\label{92w3}
\end{align}

Then there exists $u_+\in \Sigma$ such that
$$\|e^{it\Delta}u(t)-u_+\|_{\Sigma}\rightarrow 0\quad {\rm as}\quad t\rightarrow +\infty.$$
}

Obviously, Theorem 7 of this paper parallels to Theorem 2 of \cite{Song2}, the equation in (\ref{semilinear1}) only contains nonlinearities with subcritical Sobolev exponent, while the equation in (\ref{92w1}) contains nonlinearities with subcritical and critical Sobolev exponent. However, the constrictions on space dimensions and nonlinearities are different. For example, if $F(|u|^2)u=a_1|u|^{2\beta_1}u+...+a_m|u|^{2\beta_m}u$, $a_j<0$, $j=1,2,...,m$, we can take $$\frac{2-N+\sqrt{N^2+12N+4}}{4N}<\beta_1<...<\beta_m<\frac{2^*}{N}$$
and $N\geq 1$ in Theorem 7 of this paper, while we have to require that $\frac{2}{N}<\beta_1...<\beta_m<\frac{2^*}{N}$ and $N\geq 3$ in Theorem 2 of \cite{Song2}. That is, each $\beta_j$ can be smaller than $\frac{2}{N}$ or larger than $\frac{2}{N}$ in Theorem 7 of this paper, but every $\beta_j$ must be larger than $\frac{2}{N}$ in Theorem 2 of \cite{Song2}.

{\bf Remark 6.5.} If $V(x)\neq 0$ and it doesn't belong to the Kato class or $h(s)\neq 0$ for $s\geq 0$, we cannot establish classic scattering theory in this paper, we need to consider modified scattering theory because we need the results on dispersive estimates for Schr\"{o}dinger operator $-\Delta+V(x)$(see \cite{Goldberg1, Goldberg2, Schlag1, Schlag2} and the references therein). In fact, modified scattering theory is an interesting topic in the study of Schr\"{o}dinger equation with potential, we will consider it in our forthcoming paper. About the results on modified scattering theory for semilinear Schr\"{o}dinger equation with potential, we can refer to \cite{Carles03, Carles, Costin, Killip, Lu, Zhang} and the references therein.

\end{document}